\documentclass[a4paper,11pt]{article}
\pdfoutput=1 

\usepackage{jheppub} 
\usepackage{amsmath,amssymb,amsthm,amscd,graphicx,mathtools}

\input epsf.sty

\makeatletter
\newsavebox{\@brx}
\newcommand{\llangle}[1][]{\savebox{\@brx}{\(\m@th{#1\langle}\)}%
  \mathopen{\copy\@brx\kern-0.5\wd\@brx\usebox{\@brx}}}
\newcommand{\rrangle}[1][]{\savebox{\@brx}{\(\m@th{#1\rangle}\)}%
  \mathclose{\copy\@brx\kern-0.5\wd\@brx\usebox{\@brx}}}
\makeatother

\theoremstyle{definition}

\newcommand{\CC}{{\cal C}}

\newcommand{\CK}{{\cal K}}

\newcommand{\CM}{{\cal M}}

\newcommand{\CO}{{\cal O}}
\newcommand{\CP}{{\cal P}}

\def\IZ{{\mathbb Z}}
\def\IR{{\mathbb R}}
\def\IC{{\mathbb C}}
\def\IP{{\mathbb P}}

\def\IF{{\mathbb F}}

\newcommand{\re}{{\rm e}}
\newcommand{\ri}{{\rm i}}
\newcommand{\rd}{{\rm d}}

\newcommand{\mP}{\mathsf{P}}

\newcommand{\mx}{\mathsf{x}}
\newcommand{\my}{\mathsf{y}}

\newcommand{\mm}{\mathsf{p}}
\newcommand{\im}{\mathsf{i}}

\newcommand{\mO}{\mathsf{O}}
\newcommand{\mA}{\mathsf{A}}
\newcommand{\mB}{\mathsf{B}}

\newcommand{\mJ}{\mathsf{J}}

\def\({\left(}
\def\){\right)}

\newcommand{\be}{\begin{equation}}
\newcommand{\ee}{\end{equation}}
\newcommand{\ba}{\begin{aligned}}
\newcommand{\ea}{\end{aligned}}
\newcommand{\ben}{\begin{eqnarray}\displaystyle}
\newcommand{\een}{\end{eqnarray}}

\newcommand{\sectiono}[1]{\section{#1}\setcounter{equation}{0}}

\newdimen\tableauside\tableauside=1.0ex
\newdimen\tableaurule\tableaurule=0.4pt
\newdimen\tableaustep
\def\phantomhrule#1{\hbox{\vbox to0pt{\hrule height\tableaurule width#1\vss}}}
\def\phantomvrule#1{\vbox{\hbox to0pt{\vrule width\tableaurule height#1\hss}}}
\def\sqr{\vbox{%
  \phantomhrule\tableaustep
  \hbox{\phantomvrule\tableaustep\kern\tableaustep\phantomvrule\tableaustep}%
  \hbox{\vbox{\phantomhrule\tableauside}\kern-\tableaurule}}}
\def\squares#1{\hbox{\count0=#1\noindent\loop\sqr
  \advance\count0 by-1 \ifnum\count0>0\repeat}}
\def\tableau#1{\vcenter{\offinterlineskip
  \tableaustep=\tableauside\advance\tableaustep by-\tableaurule
  \kern\normallineskip\hbox
    {\kern\normallineskip\vbox
      {\gettableau#1 0 }%
     \kern\normallineskip\kern\tableaurule}%
  \kern\normallineskip\kern\tableaurule}}
\def\gettableau#1{\ifnum#1=0\let\next=\null\else
\squares{#1}\let\next=\gettableau\fi\next}

\newcommand{\figref}[1]{Fig.~\protect\ref{#1}}

\title{\boldmath Wavefunctions, integrability, and open strings}

\author{Marcos Mari\~no and Szabolcs Zakany}

\affiliation{D\'epartement de Physique Th\'eorique et Section de Math\'ematiques\\
Universit\'e de Gen\`eve, Gen\`eve, CH-1211 Switzerland}

\emailAdd{marcos.marino@unige.ch, szabolcs.zakany@unige.ch} 

\abstract{It has been recently conjectured that the exact eigenfunctions of quantum mirror curves can be obtained by combining their WKB expansion with 
the open topological string wavefunction. In this paper we give further evidence for this conjecture. We present closed expressions for the wavefunctions in the so-called maximally 
supersymmetric case, in various geometries. In the higher genus case, our conjecture provides a solution to the quantum Baxter equation of the corresponding cluster integrable system, 
and we argue that the quantization conditions of the integrable system follow from imposing appropriate asymptotic conditions on the wavefunction. We also present checks 
of the conjecture for general values of the Planck constant.}

\begin{document}

\maketitle
\flushbottom

\sectiono{Introduction}

There is by now strong evidence that topological strings on toric Calabi--Yau (CY) manifolds are closely related to spectral problems in one dimension, obtained 
by an appropriate quantization of the mirror curve. Building on previous insights in topological string theory \cite{adkmv,acdkv,km,hw}, supersymmetric 
gauge theory \cite{ns,mirmor}, and on developments in ABJM theory \cite{mp,hmo,hmo2,calvo-m, hmo3,hmmo,cgm8}, a precise formulation of this correspondence was put forward in \cite{ghm,cgm} (see \cite{mmrev} for a review). The construction developed in \cite{ghm, cgm} 
associates a set of trace class operators to a given mirror curve. Exact quantization conditions and Fredholm determinants for these operators are then conjecturally encoded in the enumerative 
geometry of the CY. This provides a correspondence between spectral theory and topological strings, or TS/ST correspondence, which has been further developed in 
\cite{kasm,mz,hatsuda-spectral, kmz,wzh,gkmr, hatsuda-comments, lst, hm, fhm, oz, bgt, grassi, swh, butterfly,sugimoto, cgum,grassi-gu,cms,ghkk,hsx,bst-2}. So far, in spite of very stringent tests, no counterexample has been found for the conjectures put forward in \cite{ghm,cgm}. 

Most of the work which has been done on the TS/ST correspondence focuses on its closed string side, which relates closed string invariants to the eigenvalue spectrum of the operators. 
However, in order to fully solve the spectral problem, one should also find the eigenfunctions. From the point of view of the TS/ST correspondence, 
this involves the open string sector. In fact, in the works \cite{adkmv,acdkv}, the central object is the D-brane wavefunction, which is the generating functional of 
certain open BPS invariants. 

A detailed study of wavefunctions in the TS/ST correspondence was made in \cite{mz-open}, focusing for simplicity on the local $\IF_0$ geometry (see \cite{amir,sciarappa} for 
other attempts to write down the wavefunctions\footnote{While this paper was being typed, a very interesting paper appeared \cite{antonio-ta} which makes a concrete proposal 
for the eigenfunctions of the relativistic Toda lattice by using instanton partition functions in the presence of defects. 
This corresponds to the family of toric CYs engineering pure $SU(N)$ gauge theories.}). Building on calculations performed in 
different limits, \cite{mz-open} conjectured that the exact wavefunctions of the spectral problem can be obtained 
by combining the WKB solution for the wavefunction with the so-called topological string wavefunction (which encodes open BPS invariants associated to symmetric Young tableaux). This is a direct 
extension of the exact results of \cite{ghm} for the spectral determinant, 
in which one combines the WKB grand potential with the topological string free energy. The conjectural wavefunctions of \cite{mz-open} 
are quantum generalizations of the Baker--Akhiezer function on the mirror curve, akin to (but different from) the construction of \cite{be}. 
However, there is a new twist in the story: as shown in \cite{mz-open}, one has also to consider different copies of 
the resulting wavefunction, corresponding to the different sheets of the Riemann surface. In the (hyper)elliptic example considered in \cite{mz-open}, 
the contribution of one of the two sheets can be easily calculated from the open BPS invariants, and then one applies an appropriate 
transformation to obtain the contribution of the second sheet. The total wavefunction is the sum of both contributions. 
Each contribution is afflicted with WKB-type singularities, which cancel in the sum. This prescription is conceptually similar to the 
mechanism described in \cite{mmss} in the context of non-critical strings. The 
total wavefunction can be written down very explicitly in the so-called maximally 
supersymmetric case or self-dual case, when $\hbar=2 \pi$. The result of \cite{mz-open} for the local $\IF_0$ geometry has been 
verified by Kashaev and Sergeev in \cite{ks}. 

The conjecture put forward in \cite{mz-open} was only developed in detail in the case of local $\IF_0$ (and for a fixed value of its mass parameter), since this is 
the simplest and most symmetric example. A deeper understanding of the open string sector for the TS/ST correspondence requires 
further testing of the conjecture in 
\cite{mz-open}. In this paper 
we make various steps in this direction, by extending the results of \cite{mz-open} in various ways. First, we test the conjecture in the maximally 
supersymmetric case $\hbar=2 \pi$ for two different geometries: 
local $\IP^2$, which has genus one, and more importantly, the resolved $\IC^3/\IZ_5$ orbifold studied in \cite{cgm}. This is a genus two geometry, which is technically 
more challenging. We manage however 
to obtain an exact expression for the wavefunctions on this genus two geometry, in the self-dual case, and for generic moduli. This result, as well as the conjectural result for 
local $\IP^2$, have been successfully checked against numerical calculations 
of the wavefunctions.

Our explicit result for a higher genus geometry allows us to explore under a new angle the relation between the quantization of mirror curves put forward in \cite{cgm}, and the 
cluster integrable system of Goncharov and Kenyon \cite{gk}. As it turns out, a toric CY leads to two different spectral problems: the spectral problem in one dimension considered in 
\cite{cgm}, based on the quantization of the mirror curve $\Sigma$, and 
the spectral problem in $g_\Sigma$ dimensions considered in \cite{gk}, based on $g_\Sigma$ mutually commuting Hamiltonians (here, $g_\Sigma$ is the genus of the mirror curve). 
The two spectral problems are however closely related. Based on the conjectural exact solution for the spectrum of the cluster integrable system proposed in \cite{fhm}, it has been noted that the spectral problem of \cite{cgm} is more general than the one associated to the cluster integrable system. In particular, the quantization condition put forward in \cite{cgm} leads to a codimension one submanifold of the moduli space. This submanifold contains the spectrum of the integrable system, which is a discrete set of points, as a 
subset \cite{fhm, swh,cgm}. It is then important to ask what is the physical mechanism which further restricts the submanifold \cite{cgm} to the discrete spectrum 
of the cluster integrable system. 
A natural answer is that the one-dimensional operator of \cite{cgm} is the Baxter operator for the cluster integrable system. 
The spectrum of the cluster integrable system should follow from 
the spectrum of the Baxter operator by requiring appropriate boundary conditions on its solutions, as it happens in the standard Toda lattice 
\cite{gutzwiller, gp, kl}. In this paper we give some evidence 
that this is the case in the example of the resolved $\IC^3/\IZ_5$ orbifold. Namely, we show that the wavefunctions of the Baxter operator, which we find explicitly 
when $\hbar=2 \pi$, decay more rapidly at infinity precisely when the values of the moduli correspond to the spectrum of the cluster integrable system. This provides a physical 
realization of the additional quantization conditions found in \cite{fhm,swh}. 

Finally, we explore the validity of the conjecture in \cite{mz-open} when $\hbar$ takes arbitrary values. In this case, 
the information provided by the open topological string amplitudes is in principle more limited: the generating functions of BPS invariants are given by 
expansions at large $x$, so we do not have closed formulae for the $x$ dependence on the wavefunctions. This leads to important limitations in obtaining
the contributions to the wavefunction from the different sheets of the Riemann surface. However, as noted in \cite{mz-open}, when the Riemann surface is hyperelliptic, the contributions of the two 
Riemann sheets can be calculated separately on the spectral theory side. It is then possible to compare the results for the contribution 
of the first Riemann sheet, which can be obtained from 
standard open BPS invariants, and we do so in the example of local $\IP^1 \times \IP^1$ and for different values of $\hbar$. We find perfect agreement.

This paper is organized as follows. In section \ref{sect-2} we present the conjecture of \cite{mz-open} for the exact eigenfunctions in a general setting, we work out in detail the 
maximally supersymmetric or self-dual case, and we illustrate it with a new example, namely local $\IP^2$. In section \ref{sect-3}, we study the genus two 
example of the resolved $\IC^3/\IZ_5$ orbifold and we make a connection between the integrable system and the decay at infinity of the wavefunctions. In section \ref{sect-4} we 
consider the conjectural eigenfunctions for arbitrary values of $\hbar$, and we study then in detail in the example of local $\IF_0$. Finally, in section \ref{sect-5} we present some 
conclusions and open problems.

\sectiono{The exact eigenfunctions: a conjecture}

\label{sect-2}

\subsection{The closed string sector}

We will now summarize some basic ingredients of the TS/ST correspondence. We refer the reader to \cite{ghm,cgm,mmrev} for more details and extensive 
references to the background results on topological string theory and local mirror symmetry. 

Let $X$ be a toric Calabi--Yau manifold, with $g_\Sigma$ ``true" moduli denoted by $\kappa_i$, $i=1, \cdots, g_\Sigma$. It 
also has $r_\Sigma$ mass parameters, $\xi_j$, $j=1, \cdots, r_\Sigma$ \cite{hkp,kpsw}. We will denote by $n_\Sigma \equiv g_\Sigma+r_\Sigma$ the total number of 
moduli of $X$. Its mirror curve has genus $g_\Sigma$ and it is 
given by an equation of the form
\be
\label{ex-W}
W(\re^x, \re^y)=0.    
\ee
It is convenient to write this curve in a ``canonical" form, by picking up one of the geometric moduli, say $\kappa_i$, so that (\ref{ex-W}) can be written as 
\be
\CO_i(x, y) +\kappa_i=0. 
\ee
The function $\CO_i(x,y)$ is a sum of monomials of the form $\re^{ax+ by}$, with coefficients that depend on the moduli and the mass parameters. We can write 
\be
\label{coxp}
 \CO_i (x,y)+\kappa_i =\CO^{(0)}_i(x,y)+ \sum_{j =1}^{g_\Sigma} \CP_{ij} (x,y) \kappa_j, 
 \ee
where $\CP_{ii}(x,y)=1$. We can obtain an operator by Weyl quantization of the mirror curve: we promote $x$, $y$ to self-adjoint Heisenberg 
operators $\mathsf{x}$, $\mathsf{y}$ satisfying the commutation relation 
\be
[\mathsf{x}, \mathsf{y}]=\im\hbar. 
\ee
Under Weyl quantization, we have that , 
\be
\re^{a x+  b y} \rightarrow \re^{a \mx + b \my}, 
\ee
so that the function $\CO_i(x,y)$ becomes a self-adjoint operator, which will be denoted by $\mathsf{O}_i$. If the mass parameters and geometric moduli satisfy 
appropriate positivity conditions, the operator 
\be
\rho_i=\mO^{-1}_i, 
\ee
acting on $L^2(\IR)$, is of trace class in all known examples \cite{kasm,lst}. Therefore, it has a discrete spectrum of 
eigenvalues $\kappa_i^{(n)}= -\re^{-E^{(i)}_n}$, $n=0,1,2, \cdots$, with eigenfunctions 
$|\psi^{(i)}_n\rangle$, which satisfy
\be
\label{spec-pro}
\left( \mO_i + \kappa_i^{(n)} \right) |\psi^{(i)}_n \rangle=0, \qquad n=0, 1, 2, \cdots
\ee
Since there are $g_\Sigma$ canonical forms for the curve, there are $g_\Sigma$ operators $\mO_i$ that one can consider. However, these operators are related by a similarity transformation 
\be
\label{op-rel}
\mO_i+ \kappa_i =\mP_{ij}^{1/2} \left( \mO_j  + \kappa_j \right) \mP^{1/2}_{ij}, \qquad i,j=1, \cdots, g_\Sigma, 
\ee
where $\mP_{ij}$ is the operator corresponding to the monomial $\CP_{ij}$. In particular, the eigenfunctions associated to the $g_\Sigma$ operators are related as \cite{cgm}
\be
\label{ji-waves}
|\psi^{(j)}_n \rangle = \mP_{ij}^{1/2} |\psi^{(i)}_n \rangle.
\ee

The conjectures of \cite{ghm,cgm,mz-open} provide an answer for this spectral problem, based on the (refined) BPS invariants of the toric CY $X$. Therefore, in order 
to write down explicit formulae for these quantities, we have to introduce some generating functionals of BPS invariants for $X$. In doing this, we will 
mostly follow the conventions of \cite{cgum}. 
As discussed above, the CY $X$ has $g_\Sigma$ ``true moduli" denoted by $\kappa_i$, $i=1, \cdots, g_\Sigma$. We will 
introduce the associated ``chemical potentials" $\mu_i$ by
\be
\kappa_i =\re^{\mu_i}, \qquad i=1, \cdots, g_{\Sigma}. 
\ee
The true moduli and the mass parameters are encoded in the Batyrev coordinates $z_i$ defined by
\be
\label{zmu}
-\log \, z_i= \sum_{j=1}^{g_\Sigma}C_{ij} \mu_j + \sum_{k=1}^{r_\Sigma} \alpha_{ik}\log {\xi_k}, \qquad i=1, \cdots, n_\Sigma.
\ee
One can choose the Batyrev coordinates in such a way that, for $i=1, \cdots, g_\Sigma$, the $z_i$'s correspond 
to true moduli, while for $i=g_\Sigma+1, \cdots, g_\Sigma+ r_\Sigma$, they correspond to mass parameters. 
For such a choice, the non-vanishing coefficients in (\ref{zmu})
\be
\label{tmc}
C_{ij}, \quad i,j=1, \cdots, g_\Sigma,
\ee
form an invertible matrix, which agrees (up to an overall sign) with the charge matrix $C_{ij}$ appearing in \cite{kpsw}. 
The mirror map expresses the K\"ahler moduli $t_i$ of the CY in terms of the Batyrev coordinates $z_i$:
\begin{equation}\label{eq:qPeriods}
	- t_i = \log z_i + \tilde{\Pi}_i(\boldsymbol{z}) \ ,\quad i =1\ldots,n_\Sigma\ ,
\end{equation}
where $\tilde{\Pi}_i(\boldsymbol{z})$ is a power series in $z_i$. Together with \eqref{zmu}, this implies that
\begin{equation}
	t_i = \sum_{j=1}^{g_\Sigma} C_{ij} \mu_j + \sum_{k=1}^{r_\Sigma} \alpha_{ik} \log {\xi_k} + \CO(\re^{-\mu}) \ .
\end{equation}
By using the quantized mirror curve, one can promote the classical mirror map to a {\it quantum mirror map} $t_i(\hbar)$ depending on $\hbar$ \cite{acdkv}:
\begin{equation}
\label{qmmap}
- t_i(\hbar) = \log z_i + \tilde{\Pi}_i(\boldsymbol{z};\hbar) \ ,\quad i =1\ldots,n_\Sigma\ .
\end{equation}
The enumerative invariants of $X$ are encoded in various important functions. The topological string genus $g$ free energies $F_g({\bf t})$ 
encode the information about the Gromov--Witten invariants of $X$. In the so-called large radius frame, they have the 
structure
\be
\label{gzp}
\ba
F_0({\bf t})&={1\over 6}\sum_{i,j,k=1}^{n_\Sigma} a_{ijk} t_i t_j t_k  + 4\pi^2 \sum_{i=1}^{n_\Sigma} b_i^{\rm NS} t_i + \sum_{{\bf d}} N_0^{ {\bf d}} \re^{-{\bf d} \cdot {\bf t}},\\
F_1({\bf t})&= \sum_{i=1}^{n_\Sigma} b_i t_i + \sum_{{\bf d}} N_1^{ {\bf d}} \re^{-{\bf d} \cdot {\bf t}}, \\
F_g({\bf t})&= C_g+\sum_{{\bf d}} N_g^{ {\bf d}} \re^{-{\bf d} \cdot {\bf t}}, \qquad g\ge 2. 
\ea
\ee
In these formulae, $ N_g^{ {\bf d}} $ are the Gromov--Witten invariants of $X$ at genus $g$ and multi-degree ${\bf d}$. The coefficients $a_{ijk}$, $b_i$ are cubic and linear 
couplings characterizing the perturbative genus zero and genus one free energies. Finally, 
 $C_g$ is the so-called constant map contribution \cite{bcov}. The constants $b_i^{\rm NS}$ 
 usually appear in the linear term of $F^{\rm NS}(\bf t,\hbar)$ (see below, \eqref{eq:NS-inst}).
The total free energy of the topological string is the formal series, 
\be
\label{tfe}
F^{\rm WS}\left({\bf t}, g_s\right)= \sum_{g\ge 0} g_s^{2g-2} F_g({\bf t})=F^{({\rm p})}({\bf t}, g_s)+ \sum_{g\ge 0} \sum_{\bf d} N_g^{ {\bf d}} \re^{-{\bf d} \cdot {\bf t}} g_s^{2g-2},   
\ee
where
\be
F^{({\rm p})}({\bf t}, g_s)= {1\over 6 g_s^2} \sum_{i,j,k=1}^{n_\Sigma} a_{ijk} t_i t_j t_k + \sum_{i=1}^{n_\Sigma} \(b_i + \frac{4\pi^2}{g_s^2} b_i^{\rm NS}\)t_i + \sum_{g \ge 2}  C_g g_s^{2g-2} 
\ee
and $g_s$ is the topological string coupling constant. 

The sum over Gromov--Witten invariants in (\ref{tfe}) 
can be resummed order by order in $\exp(-t_i)$, at all orders in $g_s$. This resummation involves the 
Gopakumar--Vafa (GV) invariants $n^{\bf d}_g$ of $X$ \cite{gv}, and it has the structure
\be
\label{GVgf}
F^{\rm GV}\left({\bf t}, g_s\right)=\sum_{g\ge 0} \sum_{\bf d} \sum_{w=1}^\infty {1\over w} n_g^{ {\bf d}} \left(2 \sin { w g_s \over 2} \right)^{2g-2} \re^{-w {\bf d} \cdot {\bf t}}.  
\ee
Note that, as formal power series, we have
\be
\label{gv-form}
F^{\rm WS}\left({\bf t}, g_s\right)=F^{({\rm p})}({\bf t}, g_s)+F^{\rm GV}\left({\bf t}, g_s\right). 
\ee
In the case of toric CYs, the Gopakumar--Vafa invariants are special cases of the 
{\it refined BPS invariants} \cite{ikv,ckk,no}. These refined invariants depend on the degrees ${\bf d}$ and on two non-negative 
half-integers or ``spins", $j_L$, $j_R$. We will denote them by $N^{\bf d}_{j_L, j_R}$. We now define the {\it Nekrasov--Shatahsvili (NS) free energy} as
\be
\label{NS-j}
F^{\rm NS}({\bf t}, \hbar) = F_{\rm NS}^{\rm pert}({\bf t}, \hbar) + F_{\rm NS}^{\rm inst}({\bf t}, \hbar) \ ,
\ee
where
\be
\label{eq:NS-pert}
F_{\rm NS}^{\rm pert}({\bf t}, \hbar) ={1\over 6 \hbar} \sum_{i,j,k=1}^{n_\Sigma} a_{ijk} t_i t_j t_k +  \( \hbar + \frac{4\pi^2}{\hbar}\) \sum_{i=1}^{n_\Sigma} b^{\rm NS}_i t_i \ ,
\ee
and
\be
\label{eq:NS-inst}
F_{\rm NS}^{\rm inst}({\bf t}, \hbar) = \sum_{j_L, j_R} \sum_{w, {\bf d} } 
N^{{\bf d}}_{j_L, j_R}  \frac{\sin\frac{\hbar w}{2}(2j_L+1)\sin\frac{\hbar w}{2}(2j_R+1)}{2 w^2 \sin^3\frac{\hbar w}{2}} \re^{-w {\bf d}\cdot{\bf  t}} \ . 
\ee
In this equation, the coefficients $a_{ijk}$ are the same ones that appear in (\ref{gzp}). 
By expanding (\ref{NS-j}) in powers of $\hbar$, we find the NS free energies at order $n$, 
\be
\label{ns-expansion}
F^{\rm NS}({\bf t}, \hbar)=\sum_{n=0}^\infty  F^{\rm NS}_n ({\bf t}) \hbar^{2n-1}. 
\ee
The first term 
in this series, $F_0^{\rm NS}({\bf t})$, is equal to $F_0({\bf t})$, the standard genus zero free energy. 

Following \cite{hmmo}, we now define the {\it grand potential} of the CY $X$\footnote{In some papers, this is also called the modified grand potential since it does not 
agree with the grand potential of the corresponding Fermi gas. In this paper we shorten the name to grand potential {\it tout court}.}. It is the sum of two functions. The first one is
\be
\label{jm2}
\ba
\mathsf{J}^{\rm WKB}_X(\boldsymbol{\mu}, \hbar)&= \sum_{i=1}^{n_\Sigma}{t_i(\hbar) \over 2 \pi}   {\partial F^{\rm NS}({\bf t}(\hbar), \hbar) \over \partial t_i} 
+{\hbar^2 \over 2 \pi} {\partial \over \partial \hbar} \left(  {F^{\rm NS}({\bf t}(\hbar), \hbar) \over \hbar} \right) \\
&+ {2 \pi \over \hbar} \sum_{i=1}^{n_\Sigma}\(b_i+b_i^{\rm NS}\) t_i(\hbar) + A({\boldsymbol \xi}, \hbar). 
\ea
\ee
The function $A({\boldsymbol \xi}, \hbar)$ is only known in a closed form in some simple geometries. The second function is 
the ``worldsheet" grand potential, which is obtained from the generating functional (\ref{GVgf}), 
\be
\label{jws}
\mathsf{J}^{\rm WS}_X(\boldsymbol{\mu}, \hbar)=F^{\rm GV}\left( {2 \pi \over \hbar}{\bf t}(\hbar)+ \pi \ri {\bf B} , {4 \pi^2 \over \hbar} \right).
\ee
It involves a constant integer vector ${\bf B}$ (or ``B-field") which depends on the geometry under consideration. This vector satisfies the following requirement: 
for all ${\bf d}$, $j_L$ and $j_R$ such that $N^{{\bf d}}_{j_L, j_R} $ is non-vanishing, we must have
\be
\label{B-prop}
(-1)^{2j_L + 2 j_R+1}= (-1)^{{\bf B} \cdot {\bf d}}. 
\ee
The {\it total grand potential} is the sum of the above two functions, 
\be
\label{jtotal}
\mathsf{J}_{X}(\boldsymbol{\mu}, \hbar) = \mathsf{J}^{\rm WKB}_X (\boldsymbol{\mu},\hbar)+ \mathsf{J}^{\rm WS}_X 
(\boldsymbol{\mu},  \hbar).
\ee
 In practice, the total grand potential can be computed by using the (refined) topological vertex \cite{akmv,ikv}, which can be used 
to compute $F^{\rm GV}$ and $F^{\rm NS}$ by taking the standard and the NS limit of the refined topological string free energy, respectively. 

The central quantity determining the spectral properties of the operator $\mO$ is the (generalized) spectral determinant of $X$. To define it, we write the quantized mirror curve as 
\be
\mO_i + \kappa_i = \mO_i^{(0)} \left(1+ \sum_{j=1}^{g_\Sigma} \kappa_j \mA_{ij} \right). 
\ee
The spectral determinant of $X$ is given by 
\be
\label{gsd}
\Xi_X ( {\boldsymbol \kappa}; \hbar)= {\rm det} \left( 1+\sum_{j=1}^{g_\Sigma} \kappa_j \mA_{ij}  \right). 
\ee
Although in defining this operator we have singled out one particular canonical form of the mirror curve (i.e. made a particular choice of $\mO_i$), it is shown in \cite{cgm} 
that the above definition is independent of this choice, so the spectral determinant is associated to the mirror curve itself, and not to any particular parametrization of it. 
The zero locus of $\Xi_X ( {\boldsymbol \kappa}; \hbar)$ defines a codimension one submanifold $\CM$ in the $g_\Sigma$-dimensional space of ``true" moduli. 
This submanifold gives the spectrum of the operator $\mO_i$ (and of the other operators 
obtained from it by similarity transformations). For example, if we fix the values of the moduli $\kappa_j$, $j\not=i$, we find a discrete set of values of $\kappa_i$ in 
$\CM$, $\kappa_{i,n}$, $n=0, 1,2, \cdots$, 
which are identified as (minus) the eigenvalues $-\re^{E^{(i)}_n}$ of $\mO_i$ (see \cite{cgm} for a detailed discussion and illustration 
in the case of the resolved $\IC^3/\IZ_5$ orbifold). 

The main conjecture of \cite{ghm,cgm} is that the spectral determinant (\ref{gsd}) can be obtained as a Zak transform of the total grand potential of $X$, as follows
\be
\label{our-conj}
\Xi_X({\boldsymbol \kappa}; \hbar)= \sum_{ {\bf n} \in \IZ^{g_\Sigma}} \exp \left( \mathsf{J}_{X}(\boldsymbol{\mu}+2 \pi \ri  {\bf n}, \hbar) \right). 
\ee
In particular, this conjecture solves completely the problem of determining the spectrum of the operator(s) associated to the mirror curve. 

\subsection{A conjecture for the exact eigenfunctions}

The total grand potential corresponds to the closed string sector of the topological string on $X$, and it solves the problem of calculating the eigenvalues of the 
quantum mirror curve. In order to extract the exact eigenfunctions, we have to find its open string theory counterpart. 
The spectral problem (\ref{spec-pro}) has a WKB solution for the eigenfunction which is a formal power series expansion in $\hbar$, 
 \be
 \label{wkbwave}
 \psi_{\rm WKB}(x; \boldsymbol{\kappa}) = \exp \left[  \sum_{n=0}^\infty S^{\rm WKB}_n(x) (-\ri \hbar)^{n-1}\right]. 
 \ee
It turns out that this expansion can be resummed, order by order in an expansion at $x\rightarrow \infty$ and at large radius. 
When expressed in terms of flat coordinates for both the open and the closed string moduli, this resummation has the following structure. Let us introduce the vector of quantum corrected 
K\"ahler parameters, obtained from the quantum mirror map (\ref{qmmap})
\be
{\bf t}_\hbar = {\bf t}({\boldsymbol \mu},\hbar), 
\ee
and the exponentiated Planck constant, 
\be
q= \re^{\ri \hbar}. 
\ee
We will use very often the exponentiated $x$ coordinate, which plays the r\^ole of the open string modulus, 
\be
X=\re^x, 
\ee
as well as its rescaled version, 
\be
\label{Xhat}
\widehat X = \re^{x-{\bf r} \cdot {\bf t}_\hbar},
\ee
where ${\bf r}$ is a vector of rational entries which depends on the geometry. Then, the {\it open string WKB grand potential} is given by 
\be
\label{jopenwkb}
\mJ^{\rm WKB}_{\rm open}(x, \boldsymbol{\mu}, \hbar)=\log \,  \psi_{\rm WKB}(x; \boldsymbol{\kappa}) = \mJ^{\rm WKB}_{\rm pert}(x, \hbar )+ \sum_{{\bf d}, \ell,s} \sum_{k=1}^\infty D^{s}_{{\bf d}, \ell} 
 {q^{k s} \over k(1-q^k)}(-\widehat X)^{-k \ell} \re^{-k {\bf d} \cdot {\bf  t}}. 
\ee
In this equation, $\mJ^{\rm WKB}_{\rm pert}(x, \hbar)$ is a perturbative part, which is a polynomial in $x$, and $D^{s}_{{\bf d}, \ell}$ are integer invariants 
which depend on a spin $s$, a winding number $\ell$, and the multi-degrees ${\bf d}$ \cite{as,amir}. The minus sign in $\widehat X$ in this equation is due to 
the fact that, in the WKB solution, the sign 
of $X$ is the opposite one to what is required by integrality of the invariants. The total WKB grand potential is obtained 
by adding (\ref{jm2}) and (\ref{jopenwkb}), i.e. 
\be
\mJ^{\rm WKB}(x, \boldsymbol{\mu}, \hbar)=\mJ^{\rm WKB}(\boldsymbol{\mu}, \hbar)+ \mJ^{\rm WKB}_{\rm open}(x, \boldsymbol{\mu}, \hbar). 
\ee
We note that, although the closed WKB grand potential can be computed from the refined 
topological vertex in the NS limit, we have not found a clear relationship between the refined vertex 
and the generating function in (\ref{jopenwkb}). In practice, we calculate (\ref{jopenwkb}) 
directly from the WKB solution for the eigenfunction. In principle 
it should be possible to calculate it also from the instanton partition 
function with defects (see \cite{sciarappa} and references therein, and \cite{antonio-ta} for very recent progress in this direction).

As in the closed string case, the open string grand potential also has a contribution from the standard open topological string. We recall that the open topological string 
free energy of a toric CY manifold $X$ depends on a choice of Lagrangian D-brane. For each choice of Lagrangian brane, one can define {\it open BPS invariants} 
$n_{g, {\bf d}, {\boldsymbol \ell}}$ \cite{ov,lmv} which generalize the Gopakumar--Vafa invariants of the closed topological strings. They depend on a quantum 
number $g$ or ``genus," the multi-degree ${\bf d}$, and winding numbers ${\boldsymbol \ell}=(\ell_1, \cdots, \ell_h)$ of the boundaries. 
The {\it topological string wavefunction} is a particular 
case of the open string free energy, depending on a single open modulus $X$ (see \cite{mz-open} for more details on this relation). It can be written in 
terms of the open BPS invariants as \cite{lmv}
\be
\begin{aligned}
\log \psi_{\rm top} (X, {\bf t}, g_s)= & \sum_{{\bf d}} \sum_{g=0}^\infty \sum_{h=1}^\infty \sum_{{\boldsymbol \ell}} \sum_{w=1}^\infty 
 {\ri^h \over h!} n_{g, {\bf d}, {\boldsymbol \ell}} {1\over w} \left( 2\sin {w g_s \over 2} \right)^{2g-2} \\
& \qquad \qquad \times  \prod_{i=1}^h \left(2 \sin {w \ell_i g_s \over 2}  \right) {1\over \ell_1 \cdots \ell_h}  X^{-w (\ell_1+\cdots+ \ell_h)}{\rm e}^{-w {\bf d}\cdot   {\bf t}}.
\end{aligned}
\label{psi-integer}
\ee
The topological string wavefunction can be computed for example by using the topological vertex \cite{akmv}. In the topological vertex formalism, 
D-brane amplitudes are given by partition functions labelled by Young tableaux. The 
topological string wavefunction involves only tableaux with a single row. 
We now introduce the worldsheet contribution to the open string grand potential, 
\be
\mJ^{\rm WS}(x,\boldsymbol{\mu}, \hbar) =\mJ^{\rm WS}(\boldsymbol{\mu}, \hbar) +\mJ^{\rm WS}_{\rm open}(x, \boldsymbol{\mu}, \hbar). 
\ee
The first term in the r.h.s. is the worldsheet grand potential (\ref{jws}), while
\be
\label{Jpsi}
\mJ^{\rm WS}_{\rm open}(x,\boldsymbol{\mu}, \hbar)=\log  \psi_{\rm top} \left( \widehat X^{2 \pi\over \hbar}, {2 \pi \over \hbar} {\bf t}_{\hbar} + \pi \ri {\bf B}, {4 \pi^2 \over \hbar} \right). 
\ee
We will sometimes use the dual Planck constant, 
\be
\label{dualh}
\hbar_D= {4 \pi^2 \over \hbar}. 
\ee
The total, $x$-dependent grand potential is 
\be
\mJ(x,\boldsymbol{\mu}, \hbar)=\mJ^{\rm WKB}(x,\boldsymbol{\mu}, \hbar)+\mJ^{\rm WS}(x,\boldsymbol{\mu}, \hbar). 
\ee
The first term in the r.h.s. of this equation is a resummation of the WKB expansion, while the second term is a non-perturbative correction in $\hbar$ to the perturbative WKB result. 
Note that both terms have poles when $\hbar/2 \pi$ is a rational number. However, as shown in \cite{mz-open}, they cancel 
when we add both functions, provided that \cite{amir}
 \be
 \label{pole-cancel}
 (-1)^{{\bf B}\cdot {\bf d}}=(-1)^{2s}
 \ee
 for all ${\bf d}$ and $s$ such that $D^s_{ {\bf d}, \ell}\not=0$. 
 
 As we just mentioned, the open topological string wavefunction depends on a choice of Lagrangian D-brane in the geometry. What is then the right choice of D-brane 
 to solve the spectral problem? It turns out that the wavefunctions associated to different branes are related by a linear canonical transformation \cite{akv}, therefore 
 they are physically equivalent and give different representations of the same wavefunction. However, one should make a choice of the Lagrangian brane which is 
 compatible with the choice of coordinate in the wavefunction. We will see some non-trivial 
 examples of this in the genus two case of section \ref{sect-3}. 

In writing the open string grand potential we have made another implicit choice, namely a choice of sheet for the Riemann surface defining the mirror curve. For example, 
when the mirror curve is hyperelliptic, in the exponent of (\ref{wkbwave}) there is an implicit choice of sign, just as in the standard WKB method. 
We will denote the choice of sheet by a subindex $\sigma$ in the 
open grand potential. When the 
mirror curve is hyperelliptic, and there are only two sheets, we have $\sigma=\pm$. The conjecture of \cite{mz-open}, 
slightly generalized to the higher genus case, states that the wavefunction $\psi(x; \boldsymbol{\kappa})$ is given by the sum over the different sheets, 
\be
\label{psiJ}
\psi(x; \boldsymbol{\kappa}) =\sum_{\sigma} \psi_\sigma (x; \boldsymbol{\kappa}), 
\ee
where
\be
\psi_\sigma (x; \boldsymbol{\kappa})= \sum_{\boldsymbol{n} \in \IZ^{g_\Sigma}}  \exp\left[ \mJ_\sigma (x,\boldsymbol{\mu}+ 2 \pi \ri \boldsymbol{n} , \hbar) \right]. 
\ee
After summing over the different sheets, we expect to find an entire function on the 
complex plane, as pointed out in \cite{mmss} in the context of non-critical strings, and as illustrated in \cite{mz-open} 
in the case of local $\IF_0$. 

There are various observations that can be made on (\ref{psiJ}). 
First of all, the wavefunction can be defined for any value of the moduli $\boldsymbol{\kappa}$.
%
However, it will not be an eigenfunction of $\rho$ unless the values of $\boldsymbol{\kappa}$ belong to the zero locus of the spectral determinant, and in many cases 
it will not even be square integrable.
For those values of $\boldsymbol{\kappa}$ where the spectral determinant vanishes, we will say that the wavefunction is ``on-shell." 
If, for example, we consider the eigenvalue equation (\ref{spec-pro}) for $i=1$ for fixed values of the moduli 
$\kappa_j$, $j=2, \cdots, g_\Sigma$, we obtain a sequence of 
eigenvalues $\kappa_1=-\re^{E_n}$. The expression (\ref{psiJ}), evaluated on these values, provides the exact 
eigenfunctions $\psi_n(x)$ corresponding to the eigenvalues. 
We can however keep the wavefunction (\ref{psiJ}) ``off-shell." In this case, the expression (\ref{psiJ}) gives an 
$x$-dependent generalization of the spectral determinant that can be calculated from the Fredholm theory of the operator $\mO$. 
This was shown in detail in \cite{mz-open} in the case of local $\IF_0$. In this and the next section, we will focus on on-shell wavefunctions, 
while in section \ref{sect-4} we will consider the theory off-shell.  

We should mention that the implementation of the sum over the different sheets turns out to be quite subtle for general values of $\hbar$. 
In the hyperelliptic case, one of the sheets (which we will take to be $\sigma=-$) involves standard BPS invariants, as obtained 
from the WKB expansion and the topological vertex. The wavefunction with $\sigma=+$ is obtained by transforming $ \psi_- (x; \boldsymbol{\kappa})$ 
to the second sheet of the Riemann surface. This can be 
done in detail in the maximally supersymmetric case, as discussed in \cite{mz-open} and in the next section, but for general values of $\hbar$ the transformation 
is more difficult to implement.

 \subsection{The maximally supersymmetric case}

One unexpected consequence of the conjectures put forward in \cite{ghm,cgm,mz-open} is that the theory becomes particularly simple when 
\be
\hbar=2 \pi. 
\ee
This is the ``self-dual" value for the Planck constant, in which $\hbar =\hbar_D$. For this value, the expressions for the spectral 
determinant and for the wavefunctions become exact at one-loop in the topological 
string expansion and in the WKB expansion. We will now write down explicit and general expressions for the wavefunctions in the maximally supersymmetric case and for 
any toric geometry. For simplicity, we will assume in the following that there are no mass parameters in the model, so the matrix $C$ reduces to 
the invertible matrix (\ref{tmc}) (the inclusion of mass parameters 
is straightforward but it requires some additional ingredients and notation).

In the self-dual case $\hbar=2 \pi$, the only contribution from the topological string wavefunction involves the disk 
amplitude $g=0, \, h=1$, and the annulus amplitude $g=0, \, h=2$. Let us introduce the functions, 
\be
\label{diska}
\ba
\widetilde D(X)&= \sum_{{\bf d}, \ell} n_{0, {\bf d}, \ell} \sum_{w=1}^\infty {1\over w^2} \re^{-w {\bf d} \cdot {\bf t} } (-\widehat X)^{- w \ell}, \\
\widetilde A(X)&=\sum_{{\bf d}, \ell_1, \ell_2} n_{0, {\bf d}, \ell_1, \ell_2 } \sum_{w=1}^\infty {1\over w} \re^{-w {\bf d} \cdot {\bf t} } (-\widehat X)^{- w (\ell_1 +\ell_2)}.
\ea
\ee
Here, we use the ``classical" K\"ahler parameters ${\bf t}\equiv {\bf t}_0$. 
Up to a change of sign in the exponentiated open string moduli, these functions are, respectively, the disk amplitude and the annulus 
amplitude $A(X_1, X_2)$ for $X_1=X_2=-\widehat X$. In order to proceed, we define two constant vectors ${\bf c}$ and ${\bf b}$ by the equality, 
\be
\label{tbc}
{\bf t}_{2\pi}+\ri \pi {\bf B} = {\bf t}({ {\boldsymbol \mu}+\ri \pi {\bf c}},0)+2\pi \ri {\bf b}. 
\ee
Using these two vectors, we can define the following transformations in the closed and open moduli, 
\be
\label{transoc}
{\boldsymbol \mu} \rightarrow {\boldsymbol \mu} +\ri \pi {\bf c}, \qquad 
	  x \rightarrow x +\ri \pi {\bf r} \cdot ({\bf B}-2 {\bf b}). 
	  \ee
We can use this transformation to obtain new functions $D(X)$, $A(X)$ from the standard disk and annulus amplitudes (\ref{diska}): 
\be
D(X)=\widetilde D(X)  \Big |_ {\tiny \begin{array}{l}
	  {\boldsymbol \mu} \rightarrow {\boldsymbol \mu} +\ri \pi {\bf c} \\
	  x \rightarrow x +\ri \pi {\bf r} \cdot ({\bf B}-2 {\bf b})
	  \end{array}}
, \qquad A(X)= 
\widetilde A(X)	  \Big |_ {\tiny \begin{array}{l}
	  {\boldsymbol \mu} \rightarrow {\boldsymbol \mu} +\ri \pi {\bf c} \\
	  x \rightarrow x +\ri \pi {\bf r} \cdot ({\bf B}-2 {\bf b})
	  \end{array}}. 
	  \ee
The remaining ingredient is the exponentially small part of the next-to-leading term in the WKB expansion, 
\be
\widetilde D_1 (X)=  \sum_{{\bf d}, \ell, s} \sum_{k=1}^{\infty} \frac{D_{{\bf d},\ell}^{s}(\frac{1}{2}-s)}{k}   \re^{-k {\bf d}\cdot {\bf t} }(-\widehat X)^{-k \ell}. 
\ee
This is essentially the one-loop correction to the WKB wavefunction. After transforming the closed and open moduli as in (\ref{transoc}), we obtain the function $D_1(X)$. 
A simple calculation by using all the above ingredients leads to the following expression
\be
\label{jminus1}
\ba
\mJ(x,\boldsymbol{\mu}, 2 \pi)&=\mJ_{\rm pert}^{\rm WKB}  (x, 2 \pi)  +{\ri \over 2 \pi} \left( x {\partial D(X) \over \partial x} + {\bf t}_{2 \pi}\cdot  {\partial D(X) \over \partial {\bf t}} - D(X)\right) \\
&-{1\over 2}  A(X) + D_1(x)+\mJ(\boldsymbol{\mu}, 2\pi). 
\ea
\ee
All the quantities appearing here can be computed explicitly in terms of geometric ingredients on the mirror curve. 
First of all, since the theory at the self-dual point $\hbar=2 \pi$ involves the 
shift of the moduli given in (\ref{transoc}), we implement this transformation directly in the equation for the mirror curve. We will denote by $y(x)$ the corresponding solution to the transformed equation. 
At large $x$, this solution goes as $y(x)=p(x)+\tilde y(x)$, where $p(x)$ is a polynomial in $x$ and $\tilde y(x)=\CO(\re^{-x})$. Let us now define the following set of differentials, 
\be
	\omega_i = -\partial_{\kappa_i} y(x) \rd x, \qquad i=1, \cdots, g_\Sigma, 
\ee 
and the associated matrix of A-periods, 
\be
	\alpha_{ij}=\oint_{\mathcal A_j} \omega_i,
\ee
which is essentially given by the derivatives of ${\bf t}_{2\pi}$ with respect to $\boldsymbol \kappa$, up to the matrix $C$ appearing in (\ref{zmu}).
By using the normalized differentials
\be
	\rd {\bf u} = \alpha^{-1} {\boldsymbol \omega},
\ee
we define the Abel-Jacobi map as 
\be 
	{\bf u}(X) =  \int_\infty ^x \rd {\bf u}, 
\ee
with the basepoint at $\infty$. A fundamental result in the open local B-model is that the disk invariants can 
be read from the equation of the mirror curve \cite{av, akv}. This leads to
\be
D(X)=  \int_{\infty}^x \tilde y(x') \rd x', \qquad \partial_{\bf t} D(X)=-2\pi \ri (C^{-1})^{\rm T} {\bf u}(X), 
\ee
where $C_{ij}$ is the matrix appearing in (\ref{zmu}). Using the above information, we can write 
\be
\label{jminus2}
\ba
\mJ (x,\boldsymbol{\mu}, 2 \pi)&=\mJ(\boldsymbol{\mu}, 2\pi) +  \mJ_{\rm pert}^{\rm WKB}(x,2\pi)+ {\ri \over 2 \pi}\Sigma( x, \boldsymbol{\mu}) -{1\over 2}  A(X) + D_1(x),  
\ea
\ee
where
\be
\label{sigmax}
\Sigma(x,{\boldsymbol \mu})=x \tilde y(x)-\int_{\infty}^x \tilde y(x')\rd x'-2\pi \ri {\bf t}_{2\pi} \cdot (C^{-1})^{\rm T}{\bf u}(X).
\ee

In order to obtain the wavefunction (\ref{psiJ}), we have to sum over all the shifts of $\boldsymbol{\mu}$ by $2 \pi \ri {\bf n}$. Only terms with explicit factors of ${\bf t}_{2\pi}$ inherit the shift:
\be
\label{n-shift}
{\bf t}_{2\pi} \rightarrow {\bf t}_{2\pi}+2\pi \ri C {\bf n}. 
\ee
To proceed, we have 
to be more explicit about the structure of the closed string contribution to the grand potential. Let us denote by 
$\widehat F_g$, $\widehat F_n^{\rm NS}$ the free energies (\ref{gzp}), (\ref{ns-expansion}) in which ${\bf t}_\hbar$ has been shifted 
by the B-field in the worldsheet instanton part. The resulting free energies have the following structure
\be
\ba
	\widehat  F_0 &= \frac{1}{6}\sum_{i,j,k=1}^{n_\Sigma} a_{ijk}t_{2\pi}^i t_{2\pi}^j t_{2\pi}^k+\widehat F_0^{\rm inst} \\
	\widehat F_1 &= \sum_{i=1}^{n_\Sigma} b_i t_{2\pi}^i+\widehat F_1^{\rm inst} \\
	\widehat F_1^{\rm NS} &= \sum_{i=1}^{n_\Sigma} b_i^{\rm NS} t_{2\pi}^i+\widehat F_1^{\rm NS, inst},
\ea
\ee
where the instanton contributions, labelled by ``inst", are invariant under the shift (\ref{n-shift}). Also, the quantity $a_{ijk}$ is totally symmetric in its labels. We then obtain
\be
\ba
\label{jshift}
	\mJ(x,{\boldsymbol \mu}+2\pi \ri {\bf n},2\pi) &= \mJ(x,{\boldsymbol \mu},2\pi) +2\ri \pi \left ( v_k+u_k(X) \right ) n_k +\ri \pi \tau_{ij} n_{i} n_{j}\\
	&-\frac{\ri \pi}{3}a_{ijk}C_{im}C_{jn}C_{kp}n_m n_n n_p, 
\ea
\ee
where repeated indices are now summed over, and 
\be
\ba
	{\bf v} &= C^{\rm T}\left [ \frac{1}{4\pi^2}\left ((\partial_{{ \bf t}_{2\pi}}^2 \widehat F_0 ){\bf t}_{2\pi}- \partial_{{ \bf t}_{2\pi}} \widehat F_0 )\right )+{\bf b}+{\bf b}^{\rm NS} \right ], \\
	\tau &= \frac{\ri}{2\pi} C^{\rm T} ( \partial^2_{{\bf t}_{2\pi}} \widehat F_0 )C.
\ea
\ee
In all the examples that have been considered, the cubic term in ${\bf} n$ in (\ref{jshift}) could always be absorbed into constant linear and quadratic terms, 
thus introducing shifts in ${\bf v} $ and $\tau$. 
We will call these shifted quantities  $\hat {\bf v} $ and $\hat \tau$. To write down the final answer for the wavefunction, we have to use the Riemann theta function with characteristics ${\bf a}$, ${\bf b}$:
\be\label{riemanntheta}
	\vartheta \begin{bmatrix*}[r]
	\mathbf a \\
	\mathbf b 
	\end{bmatrix*} 
	(\mathbf u;  \tau)
	= \sum_{\mathbf n \in \mathbb Z^{g_\Sigma}} \re^{\ri \pi (\mathbf n+\mathbf b)^{\rm T}  \tau (\mathbf n+\mathbf b) +2\ri \pi (\mathbf u+\mathbf a)^{\rm T} (\mathbf n+\mathbf b)}.
\ee
It is an odd function when $4{\bf a} \cdot {\bf b}=$odd. For definiteness, we call $\vartheta_{\rm odd}$ the theta function with ${\bf a}={\bf b}=(0,...,0,1/2)^{\rm T}$. 
The Riemann theta function with ${\bf a}={\bf b}=0$ will be denoted simply by $\vartheta({\bf u};\tau)$. The normalized B-periods of the (transformed) mirror curve 
can be written as 
\be
\oint_{\mathcal B_j} \rd {\bf u}=\tau+S,
\ee
where $S$ is a matrix of constants. According to the theory of the B-model presented in \cite{mmopen,bkmp}, 
the annulus amplitude $A(X)$ can be written in terms of the Bergman kernel of the mirror curve (see \cite{mz-open} for details of a similar computation), 
and one finds, 
\be
A(X)=\log \left ( \frac{\vartheta_{ \rm odd}({\bf u}(X);\tau+S)^2}{\mathcal C \nabla_{\bf u}\vartheta_{\rm odd}(0;\tau+S) \cdot {\bf u}'(X) } \right ) , 
\ee
where
\be
\mathcal C= \lim_{X \rightarrow \infty} X^2 \nabla_{\bf u} \vartheta_{\rm odd}(0;\tau+S) \cdot {\bf u}'(X),
\ee
is a $\kappa$ dependant constant, and ${\bf u}'(X)$ is the derivative of the Abel-Jacobi map with respect to $X$ (not $x$). Our final expression for $\psi(x; {\boldsymbol \kappa})$ is then, 
\be
\label{finalpsi}
\ba
	\psi(x; {\boldsymbol \kappa})
		& = \re^{ J ({\boldsymbol \mu},2\pi)  } \sqrt{\mathcal C \nabla_{\bf u} \vartheta_{\rm odd}(0;\tau+S) \cdot {\bf u}'(X)} \frac{ \vartheta( {\bf u}(X)+\hat {\bf v};\hat \tau)}{ \vartheta_{\rm odd}( {\bf u}(X);\tau+S) } \re^{\mJ_{\rm pert}^{\rm WKB}  (x, 2 \pi) + \frac{\ri}{2\pi}\Sigma(x,{\boldsymbol \mu}) +D_1(x)}.
\ea
\ee
This wavefunction is very similar to a classical Baker--Akhiezer function on the mirror curve \cite{akhiezer} (see for example \cite{dubrovin,bbt}), although there are also some important 
differences (for example, the term $D_1(x)$ is not part of the standard Baker--Akhiezer function). 

So far we have not been explicit about the multi-covering structure of the mirror curve. When the mirror curve is hyperelliptic, so that the 
Riemann surface is a two--sheeted covering of the complex plane, the 
wavefunction (\ref{finalpsi}) corresponds to the contribution of the first sheet $\psi_-(x; {\boldsymbol \kappa})$, and it involves the standard open BPS invariants. 
The second contribution $\psi_+(x; {\boldsymbol \kappa})$ is obtained by considering the 
transformation of (\ref{finalpsi}) to the second sheet. This involves a detailed analysis of the covering structure, but in the maximally supersymmetric case its calculation is in principle straightforward. 
Such a transformation was successfully implemented in the case of local $\IF_0$ in \cite{mz-open}, and we will see more examples in the next subsection and in section \ref{sect-3}. 
One intriguing aspect of this transformation is that the contribution of the second sheet seems to involve a 
different realization of the open string BPS invariants. We will see an illustration of this in the example of local $\IP^2$.

\subsection{An application: eigenfunctions for local $\IP^2$}

In \cite{mz-open} we used the conjecture (\ref{psiJ}) to write down an exact expression for the wavefunctions in the maximally supersymmetric case $\hbar=2\pi$ and for 
local $\IF_0$. We now apply this to another important example, namely the local $\IP^2$ geometry, also for $\hbar=2 \pi$, where we can write a fully closed expression. 

The mirror curve for local $\mathbb P^2$ is
\be
	\label{mirrorP2}
	\re^x+\re^y+\re^{-x-y}+\kappa=0.
\ee
The corresponding spectral problem is 
\be
\label{lp2-op}
\left( \mO+ \kappa\right) \psi(x)=0, \qquad \mO= \re^{\mx}+\re^{\my}+\re^{-\mx-\my}.
\ee
In order to write down the wavefunction, we have to consider the relation (\ref{tbc}). By looking at the quantum mirror map of 
local $\IP^2$ \cite{adkmv,ghm}, we find that $B=1$, $c=1$, $b=-1$. In addition, in (\ref{Xhat}) we have $r=1/3$. The transformation (\ref{transoc}) reads then, 
\be
	  \kappa \rightarrow -\kappa, \qquad 
	  x \rightarrow x +\ri \pi
\ee
We can now write down the ingredients appearing in (\ref{sigmax}). The function $\tilde y(x)$ is given by 
\be
\tilde y(X) = \log \left (\frac{-X^2-\kappa X +\sqrt{\sigma(X)}}{2 X^{-1}} \right ), 
\ee
where
\be
\sigma(X) = X(4+X(X+\kappa)^2). 
\ee
The Abel--Jacobi map is 
\be
	u(X) =\CK \frac{\partial}{\partial \kappa} \int_{\infty}^X \frac{\rd X'}{X'} \tilde y(X'),
\ee
where
\be
\CK = -\frac{3}{2\pi \ri} \left ( \frac{\partial t_{2 \pi} (\kappa)}{\partial \kappa} \right )^{-1} , \qquad 
t_{2 \pi}=3 \log (\kappa)-\frac{6}{\kappa^3} \, _4 F_3 \left ( 1,1,\frac{4}{3},\frac{5}{3};  2,2,2; \frac{27}{\kappa^3}\right ). 
\ee
The perturbative WKB piece is given by 
\be
J^{\rm WKB}_{\rm pert}(x, 2 \pi)= -{\ri x^2 \over 2 \pi}. 
\ee
For the annulus amplitude, one finds
\be
\label{annp2}
A(X)=-\log \left ( \frac{\vartheta_1(u(X);\tau)^2}{\CK^2 \vartheta_1'(0;\tau)^2}\sqrt{\sigma(X)}\right ),
\ee
where the elliptic modulus is given by 
\be
\tau=\frac{9 \ri}{2\pi} \partial_{t_{2 \pi}}^2 \widehat F_0 = \ri \sqrt{3} \,  \frac{  \, _2 F_1\left (\frac{1}{3},\frac{2}{3};1;1-\frac{27}{\kappa^3} \right ) }{  \, _2 F_1\left (\frac{1}{3},\frac{2}{3};1;\frac{27}{\kappa^3} \right ) }. 
\ee
Our conventions for the genus one theta functions are as in \cite{akhiezer}. Finally, the function $D_1(X)$ is given by 
\be
D_1(X)= {1\over 4} \log \left( {X^4 \over \sigma(X)}\right). 
\ee
This can be easily found by a standard WKB expansion. 

Using all these data, one finds, by specializing (\ref{finalpsi}), 
\be
\label{p2psi-}
	\psi_{-}(x;\kappa) = \re^{\mJ(\mu, 2 \pi)} \CK \vartheta_1'(0) \, \re^{- {\ri x^2 \over 2 \pi} +x} \frac{ \re^{ \frac{\ri}{2\pi} \Sigma(x) } }{\sqrt{\sigma(X)}} \frac{\vartheta_3(u(X)+\xi-\frac{3}{8}) }{ \vartheta_1(u(X)) },
\ee
where
\be
\ba
	 \Sigma(x)  &=  x \tilde y(X)-\int_{\infty}^X \frac{\rd X'}{X'} \tilde y(X')-\frac{2\pi \ri}{3}\, t_{2 \pi} u(X), \\
	 \xi&=\frac{3}{4\pi^2} (t_{2 \pi} \partial_{t_{2 \pi} }^2 \widehat F_0-\partial_{t_{2 \pi} } \widehat F_0 ), 
	 \ea
\ee
and the closed string grand potential $\mJ(\mu, 2 \pi)$ has been calculated in \cite{ghm}. As the subindex $-$ indicates, the expression (\ref{p2psi-}) gives just the contribution 
of the first sheet. The condition that $\psi_{-}(x;\kappa)$ decays at large $x$ is satisfied if the ratio of theta functions goes to a constant in the large $x$ limit. This happens if
\be
\label{qcp2}
	\vartheta_3 \left ( \xi-\frac{3}{8} \right )=0, 
\ee
which is precisely the quantization condition in the maximally supersymmetric case found in \cite{ghm}. This condition determines a discrete set of values for $\kappa_n =-\re^{E_n}$, 
giving the spectrum of the operator $\mO$ in (\ref{lp2-op}) when $\hbar=2 \pi$.

\begin{figure}[h]
\begin{center}
\label{numwaves}
\includegraphics[scale=0.65]{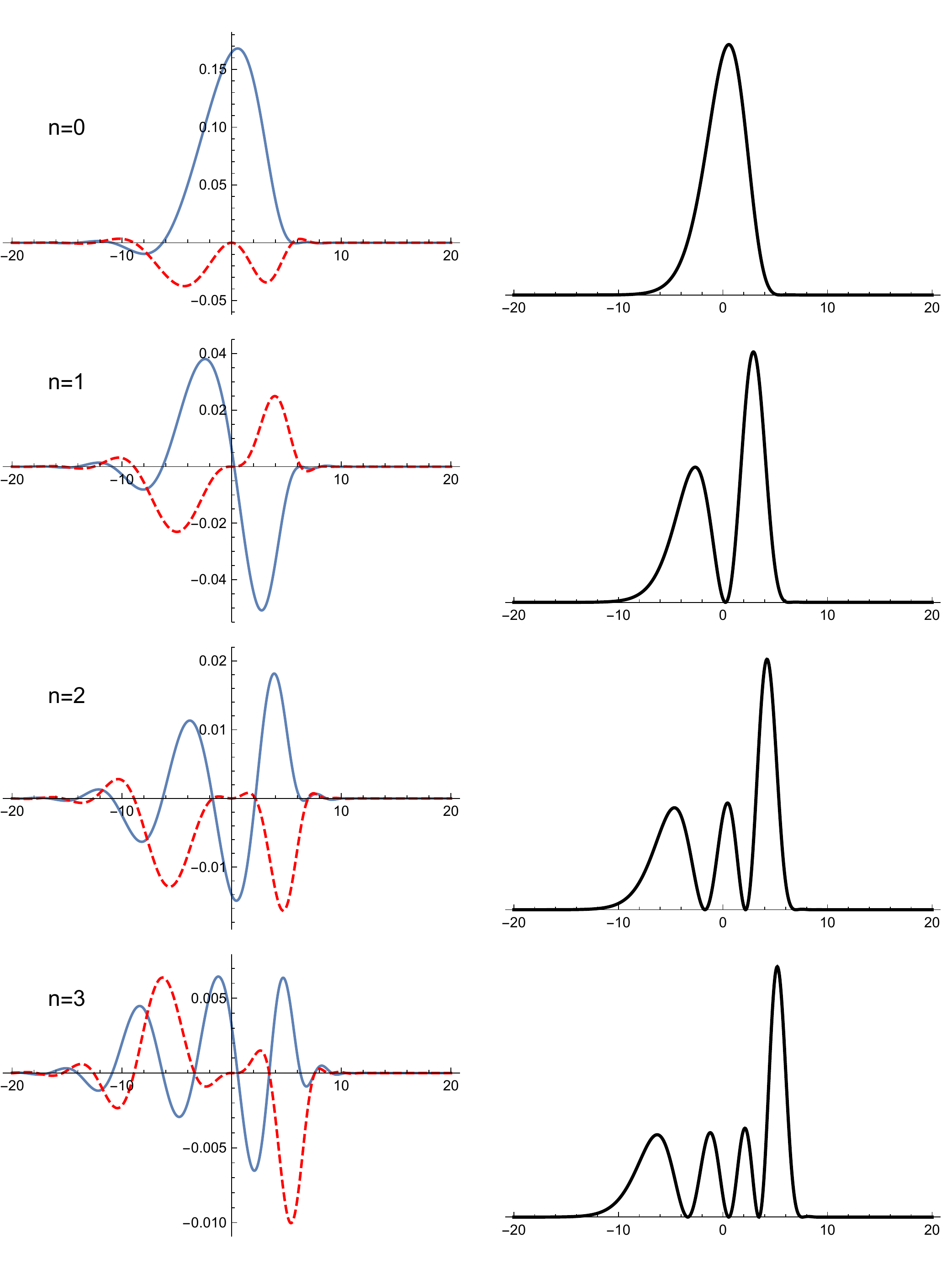}
\caption{Evaluation of the eigenfunctions of local $\IP^2$ and $\hbar=2 \pi$, by using the expression (\ref{onshellFinal}), for the ground state wavefunction and the 
first three excited states. On the left, the blue line is the real part and the red dashed line is the imaginary part. On the right we represent the square of the absolute value, showing $n+1$ peaks for the $n^{\rm th}$ level.}
\end{center}
\end{figure}

The wavefunction $\psi_{-}(x;\kappa)$ has singularities at the ``turning points" defined by $\sigma(X)=0$. In order to remove these singularities, we have to add to this function the 
wavefunction $\psi_{+}(x;\kappa)$ living in the second sheet of the Riemann surface. The transformation to the second sheet is similar to what was done in \cite{mz-open} in the 
case of local $\IF_0$. Since we want to eventually use these results to write down the actual eigenfunctions, we will assume that $\kappa = -|\kappa|+\ri 0$, with $|\kappa|>3$. 
The transformation of the Abel--Jacobi map turns out to be given by 
\be
\label{ajp2-trans}
u(X) \rightarrow -\frac{ \tau}{3}-1-u(X).
\ee
By integrating this relation and fixing the integration constant carefully, one finds 
\be
	\int_{\infty}^X \frac{\rd X'}{X'} \tilde y(X') \rightarrow-\int_{\infty}^X \frac{\rd X'}{X'} \tilde y(X')-\partial_t \widehat F_0+\frac{2\pi \ri}{3} t+{3\over  2}x^2 -\pi \ri x +{3 \pi^2 \over 2}.
\ee
In addition, the function $\tilde y(x)$ changes as 
\be
	\tilde y(x) \rightarrow 3 x - \ri \pi-\tilde y(x).
\ee
We can now write the wavefunction associated to the second sheet, 
\be
\label{p2psi+}
\psi_{+}(x;\kappa) =\re^{{ \pi \ri \over 4}} \re^{\mJ(\mu, 2 \pi)-{2 \pi \ri \over 3} \xi } \CK \vartheta_1'(0) \,  \re^{{\ri x^2 \over 4 \pi} + x} \frac{ \re^{ -\frac{\ri}{2\pi} \Sigma(x) } }{\sqrt{\sigma(X)}} \frac{\vartheta_3(u(X)+\xi-\frac{3}{8} +{\tau \over 3}) }{ \vartheta_1(u(X) +{\tau \over 3}) }. 
\ee
The total wavefunction is the sum of (\ref{p2psi-}) and (\ref{p2psi+}), and it has no singularities at the turning points. In fact, it is an entire function on the complex plane.

As in the local $\IF_0$ case analyzed in \cite{mz-open}, the 
expression for the eigenfunction simplifies considerably when one evaluates it ``on-shell," i.e. for $\kappa=-\re^{E_n}$, $n=0, 1, 2, \cdots$. This is due to the fact that, when $\xi$ satisfies 
the quantization condition (\ref{qcp2}), the quotients of theta functions in (\ref{p2psi-}) and (\ref{p2psi+}) simplify to elementary functions of $u$ and $\tau$. After some 
massaging, one finds a relatively simple formula for the eigenfunctions. To write this formula, let $X_0$ be the zero of $\sigma(X)$ given by 
\be
X_0 = \re^{x_0} = -\frac{2 \kappa}{3}-\frac{\re^{-\frac{2\ri \pi}{3}} \kappa^2}{3 \nu(\kappa)^{1/3}}+\frac{\re^{ \frac{2\ri \pi}{3}}}{3} \nu(\kappa)^{1/3},
\ee
with
\be
	\nu(\kappa)=54-\kappa^3-6\sqrt{3}\sqrt{27-\kappa^3}.
	\ee
	Let us also introduce the real K\"ahler parameter for $\kappa<0$, 
	\be
	\tilde t = 3 \log (-\kappa)-\frac{6}{\kappa^3} \, _4 F_3 \left ( 1,1,\frac{4}{3},\frac{5}{3};  2,2,2; \frac{27}{\kappa^3}\right ). 
	\ee
	Finally, we introduce the functions 
	\be
	\varphi^{\pm}_n(x) = {\rm exp} \left [\pm \frac{\ri}{2\pi} \int_{X_0}^X \rd X' \left (- \frac{\log(X')(3X'+\kappa)}{2\sqrt{\sigma(X')}}-\frac{\tilde t}{\partial_\kappa \tilde t} \frac{1}{\sqrt{\sigma(X')}} \right )  \right ]. 
	\ee
	It is understood that one should set $\kappa=\kappa_n$ in these equations. 
	Then, the eigenfunctions are given by 
	\be
	\label{onshellFinal}
	\psi_n(x; \kappa_n) = \ri { \re^{-\frac{\ri x^2}{8 \pi}+x}  \over { \sqrt{\sigma(X)}} } \left (  \varphi^+_n(x)- \varphi^-_{n}(x) \right ), 
	\ee
up to an overall normalization constant. This expression is very useful for explicit calculations. In \figref{numwaves} we show the resulting eigenfunctions for the very first energy levels, 
together with their square modulus. We have verified that these eigenfunctions agree with a direct calculation by using a standard numerical diagonalization. 

As we mentioned before, the contribution from the second sheet seems to involve a different realization of the open string invariants. This is seen more clearly in the annulus amplitude of the 
geometry. In the first sheet, this is given by (\ref{annp2}), which has the large $X$ expansion 
\be
\ba
A(X)_- &= \frac{Q+4 Q^2+35 Q^3+400 Q^4+O(Q^5)}{(-\widehat X)^2}+ \frac{2Q+6 Q^2+48 Q^3+522 Q^4+O(Q^5)}{(-\widehat X)^3} \\
	    & \quad + \frac{3Q+\frac{23}{2} Q^2+70 Q^3+690 Q^4+O(Q^5)}{(-\widehat X)^4}+\CO(\widehat X^{-5}),
\ea
\ee
where $Q=\re^{-t_{2\pi}}$.
However, after the transformation to the second sheet, implemented by (\ref{ajp2-trans}), one finds the expansion 
\be
\ba
A(X)_+ &= -\log(-\kappa X^2)+\left ( 5Q+\frac{51}{2}Q^2+\frac{806}{3}Q^3+\frac{13235}{4}Q^4+O(Q^5)
\right )\\
	    & \quad +\frac{-2+10Q^2+128Q^3+1716 Q^4+O(Q^5)}{-\widehat X}+\frac{1+3Q+4Q^2-7Q^3-325Q^4+O(Q^5)}{(-\widehat X)^2} \\
	    & \quad +\frac{-\frac{2}{3}-6Q-12 Q^2-48 Q^3-216 Q^4+O(Q^5)}{(-\widehat X)^3}
	    +\CO(\widehat X^{-4}).
\ea
\ee
Interestingly, one can also extract integer invariants from this expression by using the multicovering formula in (\ref{diska}), and they seem to correspond to a different 
open BPS sector. It would be important to have a deeper understanding 
of this new sector, associated to the second sheet of the Riemann surface. This would provide eventually a framework to obtain the precise contribution of the second sheet in the general case. 

\sectiono{Higher genus curves and integrable systems}
 \label{sect-3}
 
 As explained in \cite{cgm}, in the higher genus case, the quantization of the mirror curve \cite{cgm} leads to a 
 single quantization condition and to a codimension one submanifold $\CM$ in the space of ``true" moduli. However, the toric data of $X$ define as well a cluster integrable 
 system \cite{gk} with $g_\Sigma$ mutually commuting Hamiltonians. The spectrum of these Hamiltonians was conjecturally determined in \cite{hm,fhm} in terms of
  $g_\Sigma$ exact quantization conditions. It has been observed in \cite{fhm,swh,cgm} that the spectrum of the cluster integrable system 
 is a subspace of $\CM$. Presumably, the mechanism relating the two quantization conditions is as follows: the quantization of the mirror 
 curve gives the analogue of the Baxter operator for this problem. 
 The spectrum and eigenfunctions of this operator determine in principle the spectrum and eigenfunctions of the cluster integrable system. 
 However, there are clearly admissible eigenfunctions of the trace class operator associated to the quantum curve which are 
 not admissible solutions of the cluster integrable system, since we know that most of the points in $\CM$ are 
 not in the spectrum of the cluster integrable system. Therefore, 
 additional conditions should be imposed on the solutions of the Baxter equation. Such additional conditions 
 were empirically found in \cite{fhm} in one example, and more 
 systematically in \cite{swh}. The physical meaning of these conditions is not clear, though. 
 
 In this section, we will analyze a simple genus two geometry, namely the resolved $\IC^3/\IZ_5$ orbifold, in order to clarify this picture. 
 We will construct explicitly the eigenfunctions in the self-dual case, following 
 the prescription of the previous section. When the values of the moduli belong to $\CM$, these eigenfunctions are square integable, as expected from 
 the analysis of \cite{cgm,mz-open}. We will show however that the decay properties at infinity of these eigenfunctions change (and improve) when 
 the moduli belong to the spectrum of the corresponding cluster integrable system. In this way we will able to recover the additional quantization condition found empirically 
 in \cite{fhm}.

 \subsection{Exact wavefunctions for the resolved $\IC^3/\IZ_5$ orbifold}
  
  The resolved $\IC^3/\IZ_5$ orbifold geometry, which has $g_\Sigma=2$, 
  was studied in detail in \cite{cgm} from the point of view of the TS/ST correspondence. 
 Let us first recall some results from \cite{cgm}. The are two canonical forms for the mirror curve of this CY. The first one 
  is 
\be
	\label{firstParam}
	W_X(x',y') = \re^{x'}+\re^{y'}+\re^{-2x'-2y'}+\kappa_2 \re^{-x'-y'}+\kappa_1 = 0,
\ee
We will call this the \emph{symmetric parametrization}, because $x'$ and $y'$ appear symmetrically. The associated spectral problem is 
\be
\label{first-sp}
\left( \mO_1+ \kappa_1 \right) \psi(x')=0, \qquad \mO_1=\re^{\mx'} + \re^{\my'}+ \re^{-2 \mx'- 2 \my'} + \kappa_2  \re^{-\mx'-\my'}. 
\ee
In the second canonical form, the mirror curve is
\be
\label{secondParam}
W_X(x,y) = \re^{x}+\re^{y}+\re^{-3x-y}+\kappa_1 \re^{-x}+\kappa_2 = 0.
\ee
We will call this the \emph{hyperelliptic parametrization}, because it leads to a hyperelliptic curve in the exponentiated variables. The corresponding spectral problem is 
\be
\label{second-sp}
\left( \mO_2+ \kappa_2 \right)\psi(x)=0, \qquad \mO_2=\re^{\mx}+\re^{\my}+\re^{-3\mx-\my}+\kappa_1 \re^{-\mx}. 
\ee
The coordinates $x',y'$ and  $x,y$ appearing in (\ref{firstParam}) and (\ref{secondParam}) are related by the following linear canonical transformation
\be
\label{lct}
\begin{pmatrix} x \\ y \end{pmatrix} = \begin{pmatrix*}[r]-1 & \,\, -1 \\ 2 & \,\, 1 \end{pmatrix*}  \begin{pmatrix} x' \\ y' \end{pmatrix}.
\ee
We will focus on the hyperelliptic parametrization, since it leads to a two-sheet covering of the complex plane where we can use the simple prescriptions of the previous section. We can always 
obtain the wavefunctions in the symmetric parametrization by using (\ref{ji-waves}) As in the 
example of local $\IP^2$, we will focus on the maximally supersymmetric case in which $\hbar=2 \pi$, where we can write down explicit, closed formulae for the eigenfunctions. 
\begin{figure}[t]
\begin{center}
\includegraphics[scale=0.60]{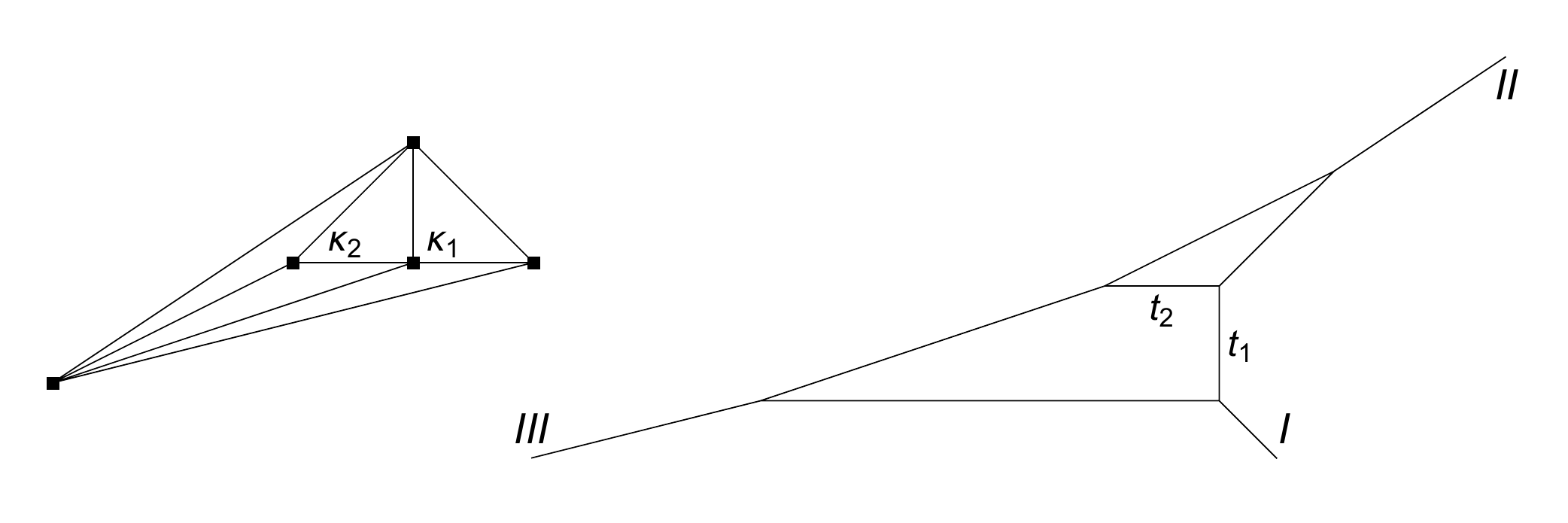}
\caption{Toric diagram and dual web for $\mathbb C^3/\mathbb Z_5$.}
\label{toricC3Z5}
\end{center}
\end{figure}

In order to write down these eigenfunctions, we recall some basic ingredients from the special geometry of the 
resolved $\mathbb C_3 / \mathbb Z_5$ orbifold. The toric and the web diagram of the 
geometry are shown in \figref{toricC3Z5}. This geometry has no mass parameters, and the Batyrev coordinates in moduli 
space are given by 
\be
	z_1=\frac{\kappa_2}{\kappa_1^3}, \qquad  z_2=\frac{\kappa_1}{\kappa_2^2}.
\ee
The corresponding K\"ahler parameters will be denoted by $t_1, t_2$ (explicit formulae for the classical and 
quantum mirror maps of this geometry can be found in \cite{cgm}). The $C$ matrix is 
\be
		C=
\begin{pmatrix*}[r]
3 & \,\, -1 \\
-1 & \,\, 2
\end{pmatrix*}.
\ee
The B-field 
is given by ${\bf B}=(1,0)$. We have to determine the vectors ${\bf r}$, ${\bf b}$ and ${\bf c}$ appearing in (\ref{Xhat}) and (\ref{tbc}). One finds that 
${\bf c}=(0,1)$, ${\bf b}=(1,-1)$ and ${\bf r}=(\frac{1}{5},\frac{3}{5})$, so (\ref{transoc}) reads
\be
\label{g2-transoc}
	  \kappa_1 \rightarrow \kappa_1, \quad \kappa_2 \rightarrow -\kappa_2, \quad 
	  x \rightarrow x +\ri \pi.
\ee
Correspondingly, the function $\tilde y(x)$ is given by 
\be	
\label{g2curve}
\tilde y(X) = \log \left ( \frac{X^3+\kappa_1 X+\kappa_2 X^2 + \sqrt{\sigma(X)}}{2 X^{3}} \right ), 
\ee
where
\be
\sigma(X) = 4 X+(X^3+\kappa_1 X+\kappa_2 X^2)^2.
\ee
The integral of $\tilde y(x)$ calculates (up to the transformation (\ref{g2-transoc})) the generating functional of disk invariants $\widetilde D(X)$ in (\ref{diska}), corresponding to a toric D-brane 
in the external leg $III$ shown in \figref{toricC3Z5}. The Abel--Jacobi map is
\be
u_i(X)=-\frac{1}{2\pi \ri} C_{i l} \left (\frac{\partial  t_{2 \pi} }{\partial \kappa} \right )^{-1}_{lj} \int_{\infty}^{X}\partial_{\kappa_j}  \tilde y(X') \frac{\rd X'}{X'},\qquad i=1,2. 
\ee
The perturbative WKB piece is 
\be
J_{\rm pert}^{\rm WKB}(x, 2 \pi)= {\ri x^2 \over 4 \pi}. 
\ee
The annulus amplitude is given by 
\be
A(X)=\log \left ( \frac{\re^{-\frac{\ri \pi}{4}} \vartheta_{\rm odd}( {\bf u}(X) ; \tau)^2 \sqrt{\sigma(X)} }{\mathcal C'(0) \mathcal C(X)} \right ), 
\ee
where the $\tau$ matrix is 
\be
\tau_{ij} = -\frac{1}{2\pi \ri} C_{im} C_{jn} \frac{\partial^2 \widehat F_0}{\partial t_{2 \pi, m} \partial t_{2 \pi, n}}
\ee
and the function $\CC(X)$ reads 
\be
\mathcal C(X) = \frac{1}{2\pi \ri} [\nabla_{\mathbf u} \vartheta_{\rm odd}(\mathbf 0)]^{\rm T}  C \left ( \frac{\partial  t_{2 \pi}}{\partial \kappa}\right )^{-1} \begin{pmatrix*}[r]
	1 \\
	X
	\end{pmatrix*}.
	\ee
Finally, the function $D_1(X)$ is given by 
\be
D_1(X)= \frac{1}{4} \log \left ( \frac{X^6}{\sigma(X)} \right). 
\ee
These ingredients determine the open string grand potential. The wavefunction (\ref{finalpsi}) is in this case given by
\be
	\label{firstSaddleWF}
	\psi_-(x;{\boldsymbol{\kappa}})  =\re^{\mJ(\boldsymbol \mu, 2 \pi)}\sqrt{\mathcal C'(0)} \sqrt{\frac{\mathcal C(X)}{\sigma(X)} } \,  \frac{
	\vartheta \begin{bmatrix*}[r]
	\mathbf 0 \\
	\mathbf 0 
	\end{bmatrix*} 
	(\mathbf u(X)+\mathbf v +  {\mathbf s} ; \tau)
	}{\vartheta_{\rm odd}( {\bf u}(X); \tau)} \,
	\re^{\frac{\ri x^2}{4\pi}+{3 x \over 2}+\frac{\ri}{2\pi}\Sigma(x)}, 
\ee
and it corresponds to the first sheet of the Riemann surface. 
In the expression (\ref{firstSaddleWF}), ${\bf v}$ is given by 
\be
	v_k = C_{ik} \left [ \frac{1}{4\pi^2} \left ( \frac{\partial^2 \widehat F_0}{\partial t_{2\pi,i} \partial t_{2\pi,j}}t_{2\pi,j}-\frac{\partial \widehat F_0}{\partial t_{2\pi,i}} \right ) +b_i +b_{i}^{\rm NS} \right ],
	\ee
where the vectors ${\bf b}$, ${\bf b}^{\rm NS}$ are, for this geometry \cite{cgm},
\be
 {\mathbf b} =\begin{pmatrix*}[r]
	 2/15 \\
	 3/20
	\end{pmatrix*}, \qquad  
	 {\mathbf b}^{\mathrm{NS}} =\begin{pmatrix*}[r]
	 -1/12 \\
	 -1/8
	\end{pmatrix*}, 
	\ee
 and the constant shift ${\bf s}$ is given by 
 \be
  {\mathbf s} =\begin{pmatrix*}[r]
	 1/2 \\
	 2/3
	\end{pmatrix*}
	\ee
 As in other cases, the quantization condition is obtained by requiring the function (\ref{firstSaddleWF}) to decay at infinity. At large $X=\re^x$ we have that,
\be
\ba
	\mathcal C(X) \approx X, \qquad \sigma(X) \approx X^6, \qquad u(X) \approx X^{-1}, \qquad \vartheta_{\rm odd}( {\bf u}(X); \tau) \approx X^{-1}. 
\ea
\ee
Therefore, in order for $\psi_-(x, {\boldsymbol{\kappa}})$ to vanish at infinity, we need to choose $\kappa_1$ and $\kappa_2$ in such a way that
\be
	\label{quantCond1}
	\vartheta \begin{bmatrix*}[r]
	\mathbf 0 \\
	\mathbf 0 
	\end{bmatrix*} 
	(\mathbf v +  {\mathbf s} ; \tau)=0.
\ee
For fixed $\kappa_1$, this gives a quantization condition for $-\kappa_2=\re^{E_2}$. Conversely, 
for fixed $\kappa_2$, this gives a quantization condition for $-\kappa_1=\re^{E_1}$. 
The quantization condition (\ref{quantCond1}) turns out to be equivalent to the vanishing of the 
spectral determinant $\Xi(\kappa_1, \kappa_2; 2 \pi)$, and it agrees with the quantization condition for this spectral problem found in \cite{cgm}.

\begin{figure}[t]
\begin{center}
\includegraphics[scale=0.75]{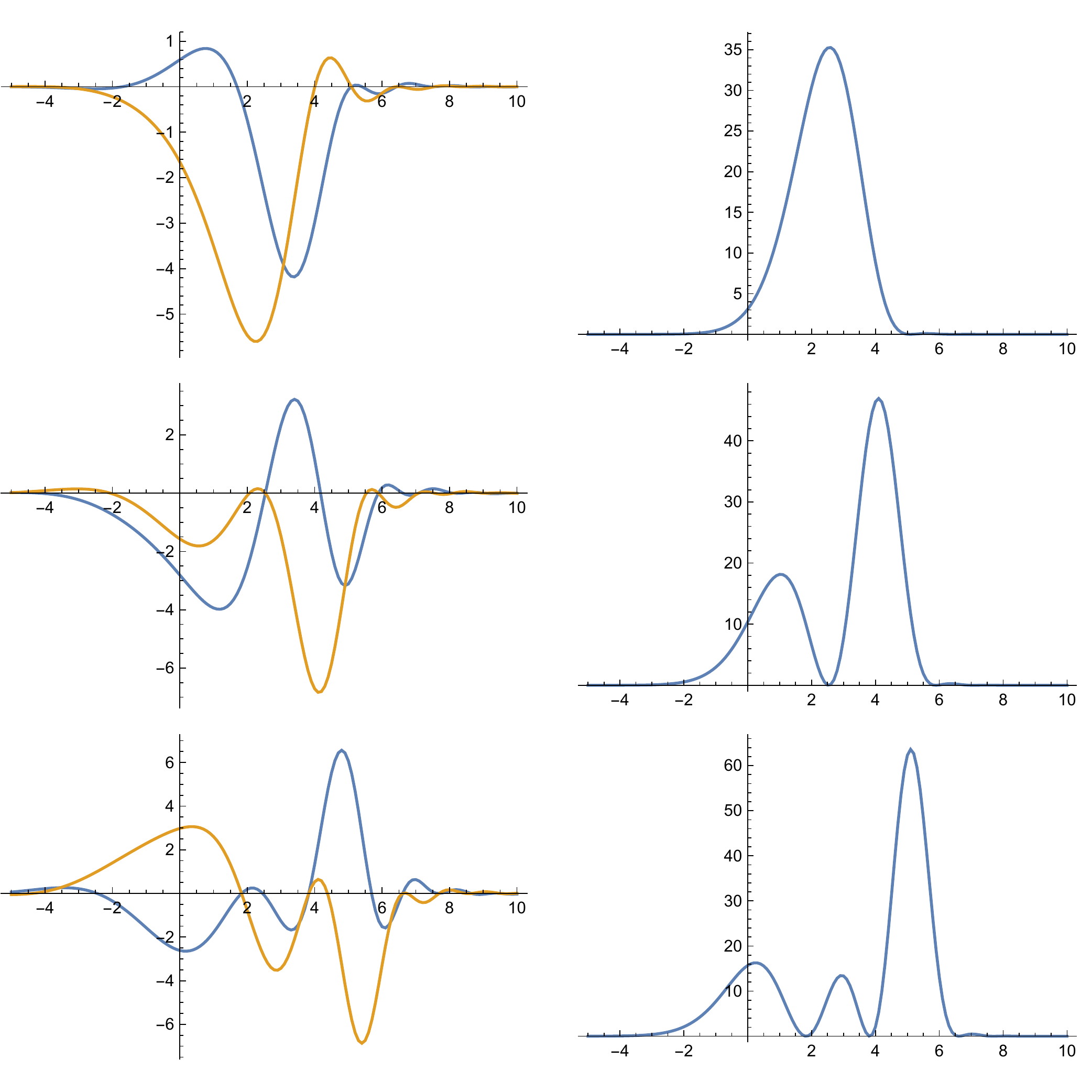}
\caption{Exact eigenfunctions for the ground state and two first excited states, as obtained numerically from (\ref{totalWF}) without the overall constants. Here we set $\kappa_1=\re^{4}$. On the left, the blue line 
shows the real part, while the orange line shows the imaginary part. The plots on the right show the squared absolute value.
\label{C3Z5wf1}}
\end{center}
\end{figure}

We should now consider the wavefunction associated to the second sheet. 
The transformation rules require a detailed analysis of the Riemann surface defined by (\ref{g2curve}). One finds that the Abel--Jacobi map 
changes as 
\be
{\bf u} (X)  \rightarrow  (\tau C^{-1}+3){\bf e}_2 - {\bf u} (X), \qquad {\bf e}_2=\begin{pmatrix*}[r]
	0 \\
	1
	\end{pmatrix*}, 
	\ee
while the integral of $\tilde y(x)$ changes as
\be
\label{gtwo2sheet}
\ba
\int_{\infty}^{X}  \tilde y(X') \frac{\rd X'}{X'} 
	\rightarrow  &
	-\frac{19\pi^2}{6}+\partial_{ t_{2 \pi, 2}} \widehat F_0-2\ri \pi\left (\frac{3}{5} t_{2 \pi, 1}+\frac{9}{5} t_{2 \pi, 2} \right )-\frac{5x^2}{2}+\ri \pi x\\
	&-\int_{\infty}^{X}  \tilde y(X') \frac{\rd X'}{X'}.
	\ea
\ee
This is valid when $\kappa_2 <0$. When $\kappa_2$ is interpreted as minus the eigenvalue of $\mO_2$, we have indeed $\kappa_2<0$ if for example $\kappa_1>0$.  
After implementing these transformations in $\psi_-(x; \boldsymbol{\kappa})$, we find
\be
\ba
	\label{secondSaddleWF}
	\psi_+(x; \boldsymbol{\kappa})
	& =\re^{\mJ(\boldsymbol \mu)}\sqrt{\mathcal C'(0)} \re^{\frac{23}{15} \pi \ri +2\pi \ri {\bf e}_2^{\rm T} C^{-1} {\bf v} } \sqrt{\frac{\mathcal C(X)}{\sigma(X)} } \,  \\
	&\qquad \qquad  \times \frac{
	\vartheta \begin{bmatrix*}[r]
	\mathbf 0 \\
	\mathbf 0 
	\end{bmatrix*} 
	(\tau C^{-1} {\bf e}_2+\mathbf v +  {\mathbf s}-\mathbf u(X) ; \tau)
	}{\vartheta_{\rm odd}(\tau C^{-1} {\bf e}_2 -{\bf u}(X); \tau)} \,
	\re^{-\frac{\ri x^2}{\pi}+{3 x \over 2}-\frac{\ri}{2\pi}\Sigma(x)}.
\ea
\ee
The full wavefunction is then the sum of (\ref{firstSaddleWF}) and (\ref{secondSaddleWF}), 
\be
\label{totalWF}
\psi(x; \boldsymbol{\kappa})=\psi_-(x; \boldsymbol{\kappa})+\psi_+(x; \boldsymbol{\kappa}). 
\ee
The resulting wavefunction is entire on the complex plane of the $x$ variable, and it belongs to $L^2(\IR)$ when the quantization condition (\ref{quantCond1}) 
is imposed. In \figref{C3Z5wf1} we show the exact eigenfunctions for the ground state and the first two excited states (we have removed the overall $x$-independent constant 
$\re^{\mJ(\boldsymbol{\mu}, 2 \pi)} {\sqrt{\CC'(0)}}$). Note that in this case $\kappa_1$ plays the r\^ole of a parameter and we have set $\kappa_1 =\re^4$. We have tested these results against 
a direct numerical calculation of the eigenfunctions, and we have found full agreement. This is a highly non-trivial test of the conjecture put forward in \cite{mz-open} in the higher genus case, where 
the solution involves full-fledged Riemann theta functions at genus two.   

Once the eigenfunctions have been found in the hyperelliptic parametrization (\ref{secondParam}), one can use the general transformation rule (\ref{ji-waves}) 
to obtain the eigenfunctions in the symmetric parametrization (\ref{firstParam}), i.e. 
for the spectral problem (\ref{first-sp}). In this case, the operator appearing in (\ref{op-rel}) is $\mP_{12}=\re^{-\mx}$. One also has to take into account 
the linear canonical transformation (\ref{lct}) relating the two variables. By implementing this transformation as a unitary operator \cite{moshinsky} (see also 
\cite{mz-open}), we find that the eigenfunction in (\ref{first-sp}) is related to the eigenfunction in (\ref{second-sp}) by 
\be
\label{wf-sym}
\psi \left ( x' \right  ) = \int  \re^{\frac{\ri}{2\hbar}(x^2-2 xx'-{x'}^2)-\frac{x}{2}} \psi(x)\, \rd x,  
\ee
up to an overall normalization constant (since our eigenfunctions are not normalized anyway, we do not keep track of these constants). 
When we plug in the integrand of the r.h.s. the eigenfunction (\ref{totalWF}) 
for parameter $\kappa_1$ and eigenvalue $-\kappa_2$, we obtain the eigenfunction of (\ref{first-sp}) with parameter $\kappa_2$ and eigenvalue $-\kappa_1$. This 
eigenfunction can be succcesfully compared 
to the result of a numerical diagonalization of the operator $\mO_1$. 

\subsection{Wavefunctions and integrability}

Armed with the explicit results obtained in the previous section, we can address now how the underlying cluster integrable system manifests itself in the 
behaviour of the eigenfunctions. First of all, we note that the eigenfunction (\ref{totalWF}), after imposing the quantization condition (\ref{quantCond1}), has the following behavior 
as $|x| \rightarrow \infty$:
\be
	\psi(x; \boldsymbol{\kappa}) \sim 
	\begin{cases}
	\re^{-x} \left ( \re^{\frac{\ri}{4\pi}x^2 } \CO(1)+  \re^{-\frac{\ri}{\pi}x^2 }\CO(1) \right ), & \qquad \qquad x \rightarrow \infty \\
	\re^{x} \left ( \re^{-\frac{3\ri}{8\pi}x^2 } \CO(1)+  \re^{-\frac{3\ri}{8\pi}x^2 }\CO(1) \right ),  & \qquad \qquad x \rightarrow -\infty.
	\end{cases}
\ee
In addition, we find that $\psi(x; \boldsymbol{\kappa})$ decays at infinity in the strip $-\frac{4\pi}{3}<{\rm Im}(x)<\frac{\pi}{2}$ around the real axis. The decay as $x\rightarrow \infty$ is guaranteed by the 
quantization condition, which can be written as
\be
\vartheta \begin{bmatrix*}[r]
	\mathbf 0 \\
	\mathbf 0 
	\end{bmatrix*} 
	({\bf u}(\infty)+\mathbf v +  {\mathbf s} ; \tau) =0, 
	\ee
to emphasize that this leads to a improved behavior when $X=\infty$. Due to Riemann's vanishing theorem, the Riemann theta function in genus two vanishes at two points on the Riemann surface. The quantization condition (\ref{quantCond1}) imposes that one of these points is $X=\infty$. In order to improve the decay properties of the wavefunction at infinity, 
we can impose the other vanishing point to be 
$X=0$, i.e. $x=-\infty$. This leads to the additional condition 
\be
\label{quantCond2}
\vartheta \begin{bmatrix*}[r]
	\mathbf 0 \\
	\mathbf 0 
	\end{bmatrix*} 
	({\bf u}(0)+\mathbf v +  {\mathbf s} ; \tau) =0.  
	\ee
When this additional condition is imposed, the decay properties of the wavefunction are enhanced to 
\be
\label{enhan}
	\psi(x; \boldsymbol{\kappa}) \sim 
	\begin{cases}
	\re^{-x} \left ( \re^{\frac{\ri}{4\pi}x^2 } \CO(1)+  \re^{-\frac{\ri}{\pi}x^2 }\CO(1) \right ), & \qquad \qquad x \rightarrow \infty \\
	\re^{\frac{3}{2}x} \left ( \re^{-\frac{3\ri}{8\pi}x^2 } \CO(1)+  \re^{-\frac{3\ri}{8\pi}x^2 }\CO(1) \right ), & \qquad \qquad x \rightarrow -\infty 
	\end{cases}.
\ee
In addition, one finds that $\psi(x)$ decays in the strip $-2\pi <{\rm Im}(x)<\frac{\pi}{2}$, which is larger than the strip obtained before. 
It can be verified that the condition (\ref{quantCond2}) 
is equivalent to the vanishing of the rotated spectral determinant considered in \cite{fhm}, i.e. to the condition 
\be
\label{rotated-xi}
\Xi(\re^{\frac{6\pi \ri}{5}} \kappa_1, \re^{-{2\pi \ri \over 5}} \kappa_2; 2 \pi)=0. 
\ee
Together, the two conditions (\ref{quantCond1}), (\ref{quantCond2}) are equivalent to the two quantization conditions proposed in \cite{fhm} to 
determine the spectrum of the cluster integrable system associated to $\IC^3/\IZ_5$.

\begin{figure}[t]
\begin{center}
\includegraphics[scale=0.6]{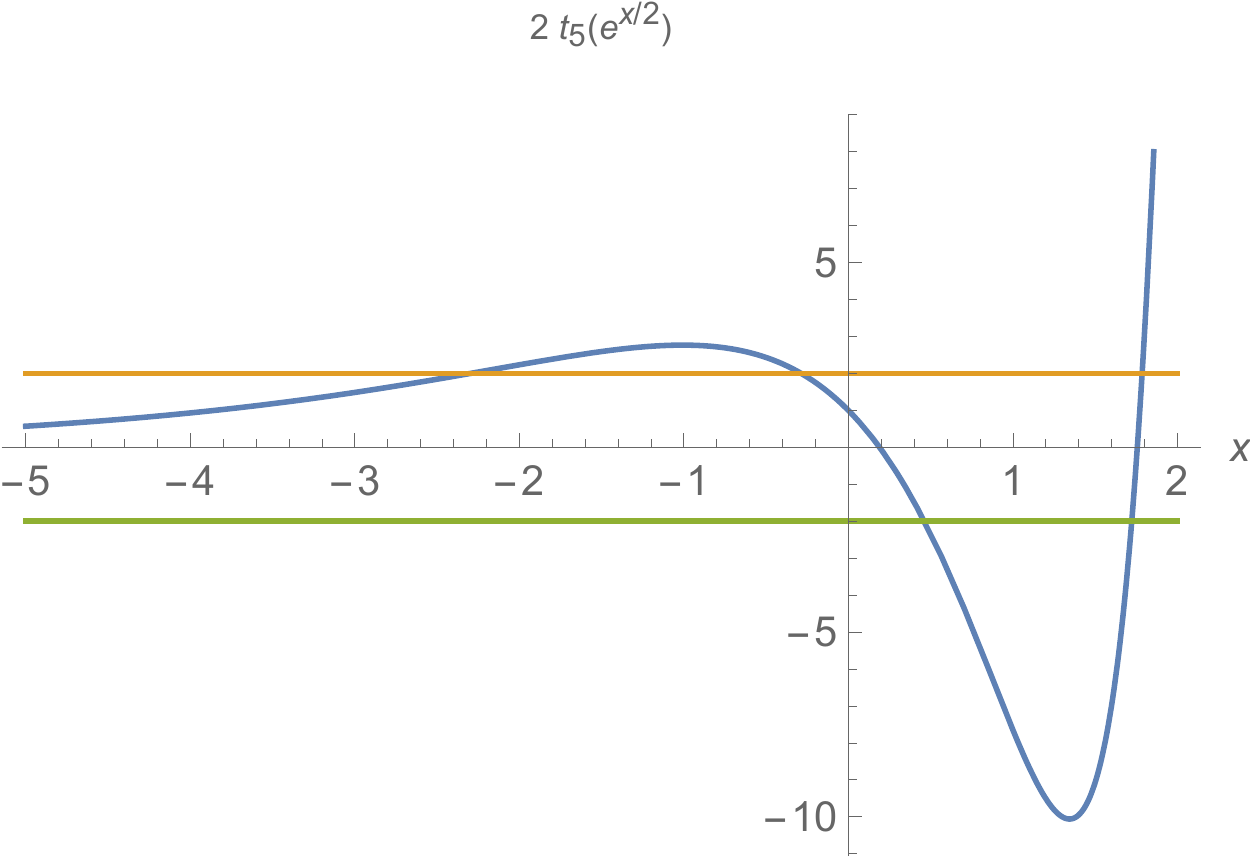}
\caption{The function $2 t_5\left(\re^{\mx/2} \right)$ as a function of $x$ and $\kappa_1=-\kappa_2=7$, showing the two ``intervals of instability" where $|t_5(\re^{x/2})|\ge 1$. 
\label{instabfig}}
\end{center}
\end{figure}

\begin{figure}[t]
\begin{center}
\includegraphics[scale=0.55]{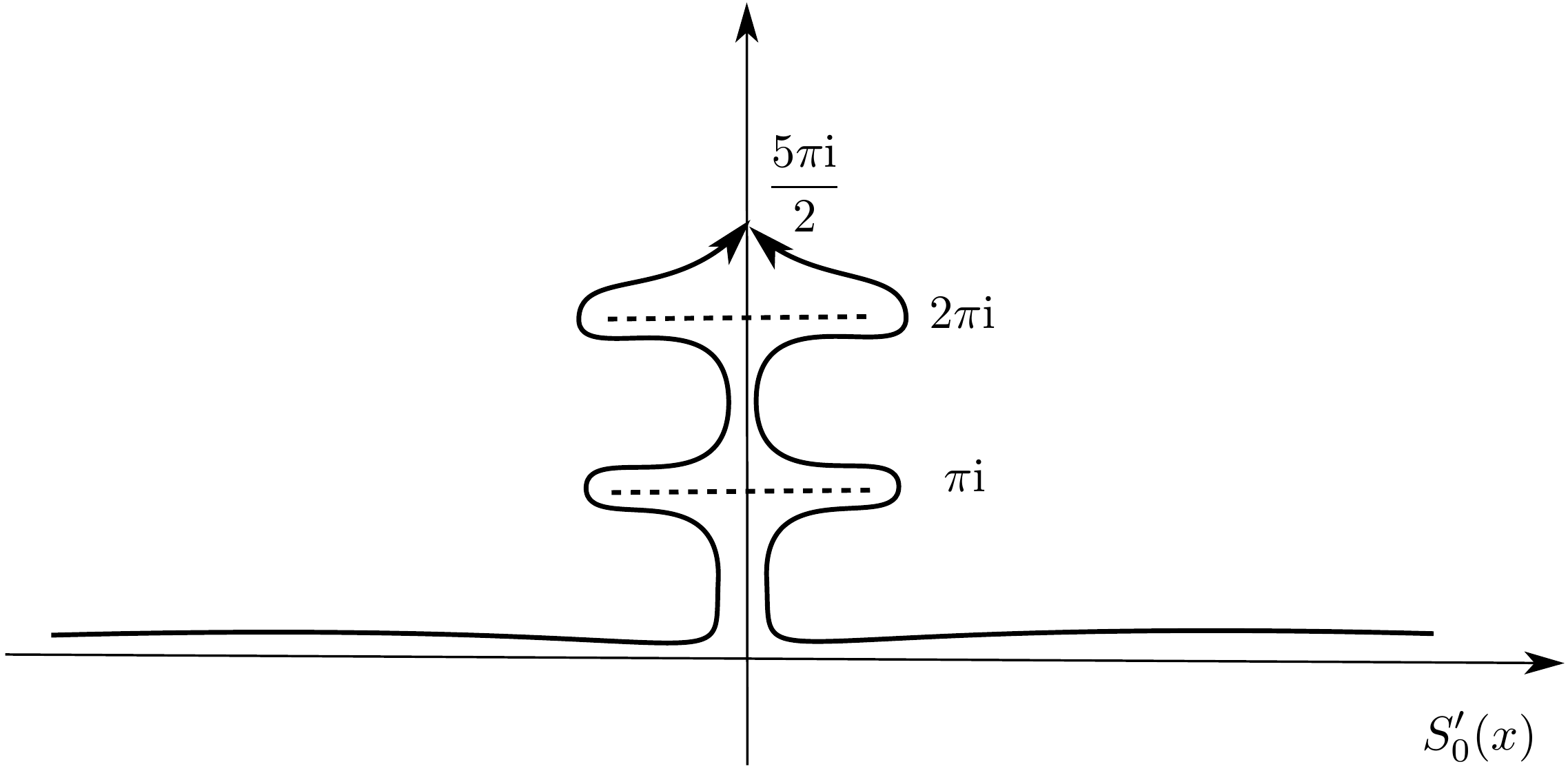}
\caption{The map $S_0'(x)$ as we go from $x\rightarrow \infty$ to $x\rightarrow -\infty$ above the real axis (line on the right) and below the real axis (line on the left). 
\label{s0fig}}
\end{center}
\end{figure}

There is a simple WKB argument which relates the quantization conditions of the cluster integrable system to the decay behavior of the eigenfunctions of 
the Baxter operator (this is based on a similar argument in \cite{gp} for the Toda lattice). We first note that, as pointed out in \cite{fhm}, under the symplectic linear transformation, 
\be
 \my \rightarrow \mm= \my + {3\over 2} \mx, \qquad \mx \rightarrow  \mx, 
\ee
the operator associated to the hyperelliptic parametrization (\ref{secondParam}) becomes
\be
\mO_2+ \kappa_2 = \re^{-{3  \mx \over 4}}\,  \mB \,  \re^{- {3  \mx \over 4}}, 
\ee
where the Baxter operator $\mB$ is given by 
\be
\mB= \re^{\mm} + \re^{-\mm} + 2 t_5(\re^{\mx/2})
\ee
and 
\be
2 t_5 (z)=z^5+\kappa_2 z^3+\kappa_1 z.
\ee
A function $Q(x)$ annihilated by the Baxter operator satisfies, 
\be
Q(x+ \ri \hbar) + Q(x-\ri \hbar) + 2 t_5(\re^{\mx/2} ) Q(x)=0, 
\ee
and the eigenfunctions of the operator $\mO_2$ are related to $Q(x)$ by 
\be
\label{relpsiq}
\psi(x; \boldsymbol{\kappa})= \re^{3x\over 4} Q(x). 
\ee
The WKB solution for $Q(x)$ is exactly of the form found in \cite{gp}, 
\be
Q(x) \approx {1\over {\sqrt{ \sinh S_0'(x)}}} \re^{-{\ri \over \hbar} \int^x S_0'(u) \rd u - {\pi x \over \hbar}}, 
\ee
where the function $S_0(x)$ is determined by 
\be
\cosh S_0'(x)= t_5(\re^{x/2}). 
\ee
Let us study the behaviour of this WKB solution as $x\rightarrow \infty$. Since 
\be
2 \cosh S_0'(x) \approx \re^{5x/2}, 
\ee
we have
\be
\label{qplus}
Q(x) \approx \re^{-{\pi x \over \hbar} -{5x \over 4}}, \qquad x\rightarrow \infty. 
\ee
As $x\rightarrow -\infty$, we have $t_5(\re^{x/2}) \rightarrow 0$, and we cross two ``intervals of instability," as shown in \figref{instabfig}. 
Between these intervals, $|t_5(\re^{x/2})|\le 1$ and $S_0'(x)$ must be imaginary. 
We can choose $S_0'(x)$ as shown in \figref{s0fig}, so that  
\be
S'_0(x) \approx {5  \pi \ri\over 2}, \qquad x\rightarrow -\infty, 
\ee
provided it satisfies the quantization conditions
\be
\label{qc-wkb}
\oint_{C_k} S_0'(u) = 2 \pi \hbar n_k, \qquad k=1,2, 
\ee
where $n_1$, $n_2$ are integers. If this is the case, $Q(x)$ behaves as
\be
\label{qminus}
Q(x) \approx \re^{3 \pi x \over 2 \hbar}, \qquad x\rightarrow -\infty. 
\ee
It is easy to verify from (\ref{qplus}), (\ref{qminus}) and (\ref{relpsiq}) that the function $\psi(x; \boldsymbol{\kappa})$ will have precisely the asymptotic behavior given in (\ref{enhan}). The 
quantization conditions (\ref{qc-wkb}) give the leading WKB approximation to the exact quantization conditions proposed in \cite{fhm}.

Our main conclusion is that, at least in this example, 
the quantization conditions of the cluster integrable system are 
conditions for an enhanced decay at infinity of the eigenfunctions of the Baxter equation. This gives a 
physical interpretation to the 
observation made in \cite{fhm}, where it was noted that 
the spectrum of the cluster integrable system is recovered when the additional condition (\ref{rotated-xi}) is imposed. It would be interesting to see 
whether the ``rotated" spectral determinants introduced in 
\cite{swh}, by generalizing the observation in \cite{fhm}, can be also related to the behavior of the eigenfunctions at infinity.

Of course, in order to have a complete picture of the relationship between the 
two spectral problems, one should find an explicit relationship between the eigenfunctions of the quantum mirror curve and the 
eigenfunctions of the cluster integrable system itself, as it was done in \cite{kl} for the Toda lattice. 
The enhanced decay properties that we have found should arise as necessary conditions for the square integrability of the eigenfunctions 
of the cluster integrable system. 

\sectiono{Testing general values of the Planck constant}
\label{sect-4}

\subsection{General strategy}

So far, our tests of the conjecture for the exact eigenfunctions have been done in the self-dual case. There is a good reason for this: when $\hbar=2 \pi$, 
one can write the functions $\psi_{\mp}(x; \boldsymbol{\kappa})$ in closed form, and in particular one can implement the transformation to the second sheet in 
complete detail, as we did in the previous examples. However, our conjecture can be also used to obtain information about the exact wavefunctions for 
general values of $\hbar$.  

In order to do this, it is useful to review some relevant aspects of the closed string case. For general $\hbar$, the total grand potential 
(\ref{jtotal}) can be computed as power series in $\re^{-{\bf t}}$, 
by using the information on the BPS invariants of $X$. From this one can in principle compute the corresponding expansion of the spectral determinant. 
This was done in some genus one geometries in section 3.2 of \cite{ghm}. It is however easier to calculate the spectral determinant by considering 
the so-called fermionic spectral traces of the operators. These are defined by the coefficients 
$ Z_X(\boldsymbol{N}; \hbar)$ in the expansion of the spectral 
determinant around the origin, 
\be
\label{or-exp}
\Xi_X ({\boldsymbol \kappa};\hbar)= \sum_{N_1\ge 0} \cdots \sum_{N_{g_\Sigma}\ge 0} Z_X(\boldsymbol{N}; \hbar) \kappa_1^{N_1} \cdots \kappa_{g_\Sigma}^{N_{g_\Sigma}}.   
\ee
This expansion can be inverted to 
\be
\label{multi-Airy}
Z_X({\boldsymbol N}; \hbar)={1\over \left( 2 \pi \ri\right)^{g_\Sigma}} \int_{-\ri \infty}^{\ri \infty} \rd \mu_1 \cdots \int_{-\ri \infty}^{\ri \infty} \rd \mu_{g_\Sigma} \,
 \exp \left\{ \mathsf{J}_X ({\boldsymbol \mu},\hbar) - \sum_{i=1}^{g_\Sigma} N_i \mu_i \right\}. 
\ee
The contour integrations along the imaginary axes can be deformed to contours where the integral is convergent. For example, in the genus one case, 
the integration contour is the one defining the Airy function, as first noted in \cite{hmo2}. It turns out that the large radius 
expansion of $\mathsf{J}_X ({\boldsymbol \mu},\hbar)$ leads to a convergent series expansion for the spectral traces, which can be 
evaluated numerically to high precision. This provides very non-trivial tests of the conjectures put forward in \cite{ghm,cgm}. 

What is the analogue of this procedure in the open string case? As already noted in \cite{mz-open}, there is an open string analogue 
of the fermionic spectral trace. This is simply obtained by expanding each of the 
wavefunctions in (\ref{psiJ}) as in (\ref{or-exp}), 
\be
\label{psin}
\psi_\sigma (x; {\boldsymbol \kappa})= \sum_{N_1\ge 0} \cdots \sum_{N_{g_\Sigma}\ge 0} \psi_{\boldsymbol{N}, \sigma}(x) \kappa_1^{N_1} \cdots \kappa_{g_\Sigma}^{N_{g_\Sigma}}.  
\ee
The analogue of the integral formula (\ref{multi-Airy}) is 
\be
\label{psi-int}
\psi_{\boldsymbol{N}, \sigma}(x) = \int_{-\ri \infty}^{\ri \infty} { \rd \mu_1 \over 2 \pi \ri} \cdots \int_{-\ri \infty}^{\ri \infty} {\rd \mu_{g_\Sigma} \over 2 \pi \ri} \,
 \exp \left\{ \mathsf{J}_\sigma (x, {\boldsymbol \mu},\hbar) - \sum_{i=1}^{g_\Sigma} N_i \mu_i \right\}. 
\ee
Note that the expansion in (\ref{psin}) 
requires that $\kappa$ takes arbitrary values. As we mentioned above, we refer to these 
as ``off-shell" wavefunctions. In \cite{mz-open} we explained how to obtain these wavefunctions 
by factorizing in an appropriate way the resolvent of the corresponding trace class operator. In this way, one can compute
the functions $\psi_{\boldsymbol{N}, \sigma}(x)$ directly in spectral theory, and for separate $\sigma$. On the other hand, the 
function $\mathsf{J}_- (x, {\boldsymbol \mu},\hbar)$ can be computed as a power series expansion 
at large radius and large open modulus $X\rightarrow \infty$, for any finite $\hbar$. By using this expansion, and integrating, 
one finds an expansion of $\psi_{\boldsymbol{N}, -}(x)$ at fixed $N$ and large $X$, where each coefficient can be computed numerically to high precision. This result 
can be then compared to the results for the off-shell wavefunctions. 

In the case of $\psi_{\boldsymbol{N}, +}(x)$, 
the calculation is more involved, since the transformation required to go to the second sheet cannot be implemented 
order by order in $1/X$ (indeed, the large $X$ expansions have different structures in different sheets, as we saw for example for the annulus amplitude in local $\IP^2$). 
For this reason, in this paper we will restrict ourselves to tests of $\psi_{\boldsymbol{N}, -}(x)$.

\subsection{The example of local $\IF_0$}

The connection between topological strings and spectral theory on local $\IF_0$ has been studied in detail in 
various references, including \cite{hw, ghm, kmz}. Studies of eigenfunctions 
have also focused on this geometry \cite{amir,sciarappa}, and it was also the main example in our previous paper \cite{mz-open}. The mirror 
curve of local $\IF_0$ is given by 
\be
\label{F0curve}
\re^x+m \re^{-x}+\re^{y}+\re^{-y}+\kappa=0.
\ee
Here, $m$ is a mass parameter that we will set to one, so that effectively we have a single K\"ahler parameter $t$. We will denote
\be
Q=\re^{-t}.
\ee
The spectral problem to be solved is, 
\be
\label{dif-f0}
(\mO +\kappa) \psi(x)=0, \qquad \mO= \re^{\mx}+ \re^{-\mx}+\re^{\my}+\re^{-\my}. 
\ee
One finds that
\be
\widehat X= Q^{1/2} X, 
\ee
and the $B$ field is in this case zero \cite{ghm}. In order to calculate $\mJ_-(x, \mu, \hbar)$ for this geometry, we first determine the WKB piece. There are two possible
methods for this. The simplest one is to solve the difference equation (\ref{dif-f0}) in a WKB expansion, order by order in $\hbar$, i.e. to calculate the functions $S_n(x)$ appearing in (\ref{wkbwave}). Then, one has to resum them in the form prescribed by (\ref{jopenwkb}). 
Another strategy consists in solving the difference equation exactly in $\hbar$, but order by order in $1/X$, akin to what was done originally in this 
example in \cite{acdkv}. Either way we obtain:  
\be
\ba
&\mJ_{\rm open}^{\rm WKB}(x, \mu, \hbar)= \mJ_{\rm pert}^{\rm WKB}(x, \hbar)\\
&\quad + \left(-\frac{q}{q-1}-\frac{2 q Q_\hbar }{q-1}+\frac{\left(q^2+q+1\right) Q_\hbar ^2}{1-q}-\frac{2
   \left(q^4+q^3+q^2+q+1\right) Q_\hbar ^3}{(q-1) q}+\CO\left(Q_\hbar ^4\right) \right){1\over {\widehat X}}\\
   &\quad + \left( \frac{q^2}{2
   \left(q^2-1\right)}+\frac{q^2 Q_\hbar }{q-1}+\frac{q \left(q^2+3 q+1\right) Q_\hbar ^2}{q^2-1}+\frac{2
   \left(q^3+q^2+q+1\right) Q_\hbar ^3}{q-1}+\CO\left(Q_\hbar ^4\right)\right){1\over {\widehat X}^2} \\
   & \quad + \CO\left(\widehat X^{-3} \right), 
   \ea
   \ee
where $Q_\hbar= \re^{-t_\hbar}$, and 
\be
 \mJ_{\rm pert}^{\rm WKB}(x, \hbar)=-\frac{\ri}{2\hbar}x^2+\frac{1}{2}\left ( \frac{2\pi}{\hbar} -1\right )x.
 \ee
The calculation of $\mJ^{\rm WS}_{\rm open}(x, \mu, \hbar)$ is even simpler, since we can use the topological vertex to resum the expansion in $\hbar_D$. We find, 
\be
\ba
\log \, \psi_{\rm top}\left(X, t, \hbar_{\rm D} \right)&=\left( \frac{\sqrt{q_D}}{q_D-1}+\frac{2 \sqrt{q_D} Q}{q_D-1}+\frac{3 \sqrt{q_D} Q^2}{q_D-1}+\frac{10 \sqrt{q_D}
   Q^3}{q_D-1} +\CO\left(Q^4\right) \right) {1\over X} \\
     & +\left( \frac{q_D}{2 \left(q_D^2-1\right)}+\frac{q_D Q}{q_D-1}+\frac{q_D (2 q_D+3)
   Q^2}{q_D^2-1}+\frac{8 q_D Q^3}{q_D-1}+\CO\left(Q^4\right)\right) {1\over X^2} \\
   &+ \CO(X^{-3}), 
   \ea
   \ee
where
\be
q_D=\re^{\ri \hbar_D}. 
\ee
These expansions can be used to calculate the open grand potential for arbitrary values of $\hbar$. The expression (\ref{psi-int}) becomes, in this genus one example, 
\be
\label{psin-simple}
\psi_{-,N}(x) = \int_{\mathcal C} \frac{\rd \mu}{2\pi \ri}\,  \re^{J(\mu,X,\hbar)-N\mu},
\ee
where $\CC$ is the integration contour for the Airy function, used in many previous computations (see for example \cite{ghm}). 
We recall that the closed string grand potential has the structure \cite{ghm}, 
\be
\mJ(\mu, \hbar) =\frac{C(\hbar)}{3}\mu^3+B(\hbar)\mu + A(\hbar) +\CO\left(\re^{-\mu} \right), 
\ee
where $A(\hbar)$, $B(\hbar)$, $C(\hbar)$ are calculable constants. We can then write the integrand in the l.h.s. of (\ref{psin-simple}) as a double expansion at large $\mu$ and $X$, 
\be
\re^{\mJ(\mu,X,\hbar)-N \mu}= \re^{  { C(\hbar) \over 3}  \mu^3-  (-B(\hbar)+N)\mu + A(\hbar) + \mJ_{\rm pert}^{\rm WKB}(x, \hbar)} 
 \sum_{\alpha,\beta} \frac{x^\beta}{X^{\alpha}} \sum_{n,\ell} c^{(\alpha,\beta)}_{\ell,n}(\hbar) 
 \re^{-n \mu} \mu^\ell, 
 \ee
We obtain, 
\be
	\label{psim-conj}
\psi_{-,N}(x) = \re^{  \mJ_{\rm pert}^{\rm WKB}(x, \hbar)}\sum_{\alpha,\beta}  f_{\alpha,\beta}(N,\hbar) \frac{x^\beta}{X^{\alpha}}, 
\ee
where the numerical coefficients $ f_{\alpha,\beta}(N,\hbar) $ are given by a (convergent) sum of Airy functions, 
\be
\label{fab}
 f_{\alpha,\beta}(N,\hbar)=\re^{A(\hbar)}  C^{-1/3}(\hbar) \sum_{n,\ell}c^{(\alpha,\beta)}_{\ell,n}(\hbar)  \left ( -\frac{\partial}{\partial N} \right )^\ell  {\rm Ai} \left ( \frac{N-B(\hbar)+n}{C^{1/3}(\hbar)} \right ). 
 \ee
 This is the prediction of our conjecture for the values of the wavefunctions $\psi_{-,N}(x)$, in terms of open and closed BPS invariants of the geometry 
 (which are encoded in the coefficients $c^{(\alpha,\beta)}_{\ell,n}(\hbar)$).

 To illustrate these predictions even more concretely, let us consider the value $\hbar=4\pi$, which is particularly useful for a comparison 
with the results of spectral theory. One finds the following double expansion of the open string grand potential, at large $\mu$ and large $X$:
\be
	\label{J4pi}
\ba
\mJ(x, \mu, 4 \pi)&= \mJ(\mu, 4 \pi) -{\ri \over 8 \pi} x^2 -{x\over 4} \\
&-\left\{  \frac{\ri}{2
  } \re^{\mu/2}+\ri \re^{-\mu/2}+\ri \re^{-3\mu/2}+6 \ri
   \re^{-5\mu/2}+\CO\left(\re^{-7\mu/2}\right)\right\}{1 \over \sqrt{X}} \\
   & + \biggl\{ \left (-\frac{\ri x}{4 \pi }+\frac{\ri \mu }{4 \pi
   }-\frac{\ri}{4 \pi }-\frac{1}{2} \right) \re^{\mu}+\frac{1}{2}+\left(-\frac{\ri \mu
   }{\pi }-\frac{\ri}{2 \pi }+1\right) \re^{-\mu}+4 \re^{-2\mu} \\
   & \qquad   +\left(-\frac{5 \ri \mu }{\pi }-\frac{\ri}{4 \pi }+12\right) \re^{-3\mu}+\CO\left(\re^{-4\mu}\right) \biggr\} {1\over X} + \CO\left( X^{-3/2} \right). 
   \ea
   \ee
The closed string grand potential can be computed with the techniques of \cite{ghm}. One finds, 
\be
\mJ(\mu, 4 \pi)= \frac{\mu ^3}{6 \pi ^2}-\frac{\mu
   }{4} +A(4\pi)-\re^{-\mu}+ \left(-\frac{2 \mu ^2}{\pi
   ^2}-\frac{\mu }{\pi ^2}-\frac{1}{2 \pi ^2}\right)\re^{-2\mu}-\frac{16}{3}\re^{-3\mu}+\CO\left(\re^{-4\mu}  \right). 
   \ee
 In spectral theory, the function $\psi_{-, N}(x) $ for $\hbar=4 \pi$ can be computed exactly as it was done in \cite{mz-open} in the self-dual case. 
 One finds, for $N=0$, the following expressions:
 \be
 \ba
 	\psi_{-,0}(x) &= \re^{ -\frac{\ri x^2}{8\pi}-\frac{x}{4}} \frac{ \re^{\frac{5 \ri \pi}{16}} \re^x \left(-\ri \sqrt{2}
   \re^{x/2}+\re^x-\ri\right)}{2 \sqrt{\pi }
   \left(\re^{2 x}-1\right)} \\
	&=\re^{  \mJ_{\rm pert}^{\rm WKB}(x, \hbar)}\frac{\re^{\frac{5 \ri \pi}{16}}}{2\sqrt{\pi}} \left (1-\ri \sqrt{2}X^{-1/2} -\ri X^{-1}+X^{-2}+...\right ).
\ea
 \ee
 For $N=1$ and $N=2$, the exact expressions are somewhat long, but their expansion reads, 
 \be
 \ba
  \psi_{-,1}(x) &=\re^{  \mJ_{\rm pert}^{\rm WKB}(x, \hbar)}\frac{\re^{\frac{5 \ri \pi}{16}}}{2\sqrt{\pi}} \left (\frac{4-\pi}{16\pi}-\frac{\ri}{8\sqrt{2}} \sqrt{2}X^{-1/2} +\frac{\ri (-4 x+(1+10 \ri) \pi +4)}{16 \pi }X^{-1}+...\right ), \\
\psi_{-,2}(x) &=\re^{  \mJ_{\rm pert}^{\rm WKB}(x, \hbar)}\frac{\re^{\frac{5 \ri \pi}{16}}}{2\sqrt{\pi}} \left (\frac{5\pi^2-8 \pi-24}{512 \pi^2}-\frac{\ri(\pi^2-8)}{256\sqrt{2}\pi^2} X^{-1/2} \right. \\
	& \qquad \qquad \qquad \qquad \left. +\frac{ 8 \ri(\pi-4) x+\pi^2(20+3\ri)-\pi(80+8\ri)+24\ri}{512 \pi^2 }X^{-1}+...\right ).
\ea
 \ee
The coefficients of the monomials $x^\beta X^{-\alpha}$ inside the parentheses are reproduced by our Airy formula (\ref{fab}) (up to the overall normalization factor 
$\re^{\frac{5 \ri \pi}{16}}/2\sqrt{\pi}$). Expanding the grand potential in (\ref{J4pi}) up to order $\re^{-3\mu}$ (as it is given in the explicit expression) yields around 16-18 significant digits. If we increase the number of terms and use an expansion up to order $\re^{-6\mu}$, the precision is increased to 30-32 significant digits. This provides a strong check of our conjecture at $\hbar=4\pi$. 

The same procedure can be performed for other values of $\hbar$, for example $\hbar=2\pi/3$. For that value, the exact $\psi_{-,N}(x)$ can also be expressed using elementary functions. The grand potential for $\hbar=2\pi/3$ is
 \be
 \ba
 	\mJ\left (x,\mu,\frac{2\pi}{3} \right ) &=\mJ\left (\mu,\frac{2\pi}{3} \right )-\frac{3\ri x^2}{4\pi}+x \\
	& + \left \{-\frac{3-\ri \sqrt{3}}{6} \re^\mu \right \}\frac{1}{X}+ \left \{\frac{(3+\ri \sqrt{3}) \re^{2\mu} }{12} -1 \right \}\frac{1}{X^2}+\left \{  -\frac{\pi+\ri(1+3x-3\mu)}{6\pi}\re^{3\mu} \right. \\
	& \left. +\frac{(9\pi+\ri \sqrt{3})\pi+18\ri(x-\mu)}{6\pi} \re^{\mu}+\CO(\re^{-\mu})\right \} \frac{1}{X^3}+\CO(X^{-4}),
 \ea
 \ee
 with $\mJ\left (\mu,\frac{2\pi}{3} \right )$ given by
 \be
 \ba
 	\mJ\left (\mu,\frac{2\pi}{3} \right ) =\frac{\mu^3}{\pi^2}+\frac{4\mu}{9}+A\left (\frac{2\pi}{3}\right )+\frac{-4\pi^2-54\mu^2+3\sqrt{3}\pi(2\mu+1)}{9\pi^2}\re^{-2\mu} +\CO(\re^{-4\mu}).
 \ea
 \ee
 Again, by using (\ref{fab}), we find perfect agreement with the exact wavefunctions computed from spectral theory.

\sectiono{Conclusions and open problems}

\label{sect-5}
In this paper we have provided various tests of the conjecture of \cite{mz-open} for the exact eigenfunctions of quantum mirror curves. 
We have verified it in the local $\IP^2$ geometry and in the resolved $\IC^3/\IZ_5$ orbifold geometry, which has genus two, for the self-dual value of the Planck constant. 
We have also tested it for local $\IF_0$, as originally done in \cite{mz-open}, but 
for more general values of $\hbar$. In all cases, we have found a remarkable agreement. Our results provide the full conjectural solution 
of very non-trivial spectral problems, as it should be clear from the example of the resolved $\IC^3/\IZ_5$ orbifold, which involves genus two Riemann theta functions. 
In addition, we have used these results to clarify the relation between the 
quantum mirror curve and the underlying cluster integrable systems: at least in the example considered in this paper, 
the spectrum of the integrable system corresponds to values of the moduli where the 
eigenfunctions have an enhanced decaying behavior at infinity. It would be very interesting to see if the additional quantization conditions introduced in \cite{swh}, by generalizing 
the observation in \cite{fhm}, can be 
also interpreted in terms of decay properties of the corresponding eigenfunctions. 

There are clearly many problems that remain open. An important ingredient of our solution is that we have to sum over different sheets of the 
Riemann surface in order to obtain the correct eigenfunction. Our conjecture (\ref{psiJ}) predicts in particular that the different sheets 
contribute equally. So far, this summation has been implemented only when the mirror curve is hyperelliptic, but one should understand more general situations. 
Even in the hyperelliptic case, the transformation rules to write the wavefunctions in different sheets can be applied in detail only in the 
self-dual case (i.e. when we have a fully resummed function 
of $x$). For general values of $\hbar$, one needs 
more work to find a concrete prescription to obtain the off-shell wavefunction on the second sheet. 
As we mentioned in this paper, the existing evidence seems to indicate that this wavefunction involves a different topological open string sector, with new integer 
invariants. Clearly, this deserves further investigation. 

A related issue is that, for general values of $\hbar$, the conjecture provides an expression for the wavefunction as an expansion 
around $x\rightarrow \infty$. It would be important to find other representations, in which the dependence on $x$ is partially resummed. In recent 
work \cite{antonio-ta}, it has been shown that the instanton partition function with defects gives the building block for such a partial resummation. 
It would be interesting to see whether this representation sheds some 
light on our conjecture. 

Another important open problem, which was already mentioned in section \ref{sect-3}, concerns 
the relation between the quantum mirror curve and the cluster integrable system. As in the case of the Toda lattice, 
we should find an explicit relationship between the eigenfunctions of the two different quantum problems\footnote{A partial resolution of this problem in the 
relativistic Toda lattice, which is relevant to our framework, was presented in \cite{klst}.}. This will probably give a deeper rationale for our 
observations on the decay properties of the eigenfunctions. 

Finally, the wavefunctions associated to the quantum mirror curves represents only a small subset of D-brane partition functions (those corresponding 
to symmetric Young tableaux). It would be important to find spectral theory implementations of more general open string amplitudes, providing in this way a non-perturbative definition 
of the full open string sector.

\section*{Acknowledgements}
We would like to thank Santiago Codesido, Rinat Kashaev and Antonio Sciarappa for useful discussions and correspondence.
This work is supported in part by the Fonds National Suisse, 
subsidies 200021-156995 and 200020-141329, and by the NCCR 51NF40-141869 ``The Mathematics of Physics" (SwissMAP). 

\appendix

\end{document}